\documentclass{article}

\usepackage{algorithm}
\usepackage[margin=30mm,includehead,includefoot]{geometry}
\usepackage[noend]{algpseudocode}
\usepackage{amsmath,amssymb,amsthm,amscd,color,comment}
\usepackage{autobreak}
\usepackage{color}
\usepackage{url}
\usepackage{blkarray}
\usepackage{jheppub_mod}
\usepackage{empheq}
\usepackage{graphicx}
\usepackage{booktabs, multirow}
\usepackage{subcaption}
\usepackage{bm}
\usepackage{enumerate}
\usepackage{microtype}
\usepackage{mathtools}
\usepackage{wrapfig}
\usepackage{caption}
\usepackage{enumitem}

\usepackage{multicol}

\usepackage{tikz}
\usetikzlibrary{patterns}
\usetikzlibrary{arrows,shapes,positioning}
\usetikzlibrary{trees}
\usetikzlibrary{matrix,arrows} 				
\usetikzlibrary{positioning}				
\usetikzlibrary{calc,through}				
\usetikzlibrary{decorations.pathreplacing}  
\usepackage{pgffor}							
\usetikzlibrary{decorations.pathmorphing}	
\usetikzlibrary{decorations.markings}
\tikzstyle{block} = [draw, rectangle, 
minimum height=3em, minimum width=6em]

\tikzstyle{decision} = [diamond, draw, fill=blue!20, 
text width=4.5em, text badly centered, node distance=3cm, inner sep=0pt]
\tikzstyle{block} = [rectangle, draw, fill=blue!10, 
text width=7em, text centered, rounded corners, minimum height=4em, node distance=3cm]
\tikzstyle{line} = [draw, -latex']
\tikzstyle{cloud} = [draw, rectangle, fill=blue!20, text width=6em, text centered, rounded corners, node distance=3cm,
minimum height=2em]
\tikzstyle{border} = [draw, dashed, rectangle, fill=blue!5, rounded corners, node distance=3cm, minimum height=25em, minimum width=22em]
\tikzstyle{data} = [draw, rectangle, fill=blue!10, rounded corners, minimum height=2em, minimum width=8em]
\tikzstyle{box} = [draw, rectangle, fill=white, rounded corners, minimum width=10em]
\tikzstyle{input}=[trapezium, draw, text centered, trapezium left angle=60, trapezium right angle=120, minimum height=2em, fill=blue!10]

\tikzset{fermionnoarrow/.style={draw=black},	
        source/.style={draw=black, postaction={decorate}},
	    rgrav/.style={decorate, decoration={snake}, draw},
        bgrav/.style={dashed,dash pattern=on 3pt off 1.5pt,draw=black, line width=1pt, postaction={decorate}},
        cross/.pic = {
    \draw[rotate = 45] (-#1,0) -- (#1,0);
    \draw[rotate = 45] (0,-#1) -- (0, #1);
    }
     }

\definecolor{green1}{HTML}{3D792A}
\definecolor{cyan1}{HTML}{37cdaa}
\definecolor{blue1}{HTML}{5d7ac4}
\definecolor{red1}{HTML}{d0482a}
\definecolor{purple1}{HTML}{845ea8}
\definecolor{orange1}{HTML}{e07229}

\title{
Renormalizing Love:~tidal effects at the third post-Newtonian order
}

\author[a,b,c]{Manoj K. Mandal,}
\author[a,b]{Pierpaolo Mastrolia,}
\author[d]{Hector O. Silva,}
\author[d,e]{Raj Patil,}
\author[d]{Jan Steinhoff}

\newcommand{\unipd}{Dipartimento di Fisica e Astronomia, Universit\`a degli Studi di Padova,
Via Marzolo 8, I-35131 Padova, Italy.}

\newcommand{\pdinfn}{INFN, Sezione di Padova,
Via Marzolo 8, I-35131 Padova, Italy.}

\newcommand{\ucla}{Mani L. Bhaumik Institute for Theoretical
Physics, University of California at Los Angeles, Los Angeles, CA
90095, USA.}

\affiliation[a]{\unipd}
\affiliation[b]{\pdinfn}
\affiliation[c]{\ucla}
\affiliation[d]{Max Planck Institute for Gravitational Physics (Albert Einstein Institute), Am M{\"u}hlenberg 1, Potsdam D-14476, Germany}

\affiliation[e]{Institut f{\"u}r Physik und IRIS Adlershof, Humboldt-Universit {\"a}t zu Berlin, Zum Großen Windkanal 2, D-12489 Berlin, Germany}

\emailAdd{manojkumar.mandal@pd.infn.it}
\emailAdd{pierpaolo.mastrolia@unipd.it}
\emailAdd{hector.silva@aei.mpg.de}
\emailAdd{raj.patil@aei.mpg.de}
\emailAdd{jan.steinhoff@aei.mpg.de}

\abstract{ 
We present the conservative effective two-body Hamiltonian at the third order in the post-Newtonian expansion with gravitoelectric quadrupolar dynamical tidal-interactions.
Our derivation of the effective two-body Lagrangian is based on the diagrammatic effective field theory approach and it involves Feynman integrals up to three loops, which are evaluated within the dimensional regularization scheme.
The elimination of the divergent terms occurring in the effective Lagrangian requires the addition of counterterms to ensure finite observables, thereby introducing a renormalization group flow to the post-adiabatic Love number.
As a limiting case of the renormalized dynamical effective Hamiltonian, we also derive the effective Hamiltonian for adiabatic tides, and, in this regime, calculate the binding energy for a circular orbit, and the scattering angle in a hyperbolic scattering.
}

\def\NNLO{N$^2$LO }
\def\NNNLO{N$^3$LO }

\newcommand{\dd}{{\rm d}}
\newcommand{\DD}{{\rm D}}
\newcommand{\ii}{{\rm i}}

\allowdisplaybreaks

\begin{document}
\addtocontents{toc}{\protect\setcounter{tocdepth}{2}}

\begin{flushright}
\begingroup\footnotesize\ttfamily
	HU-EP-23/43-RTG
\endgroup
\end{flushright}

\maketitle

\section{Introduction}

A new era in astronomy and cosmology has begun with the recent gravitational wave (GW) detections by the LIGO-Virgo-KAGRA collaboration \cite{LIGOScientific:2021djp}. The worldwide network of ground-based \cite{LIGOScientific:2014pky,VIRGO:2014yos,KAGRA:2020agh,Saleem:2021iwi,LIGOScientific:2016wof,Punturo:2010zza}, as well as space-based GW detectors \cite{LISA:2017pwj} continues to grow, and will grant access to an ever broader frequency band with higher sensitivity. 

Among the most significant sources of GWs are the neutron star (NS) binaries~\cite{LIGOScientific:2017vwq,LIGOScientific:2017ync,LIGOScientific:2020aai}, which provide insights into the physics of dense nuclear matter within these stars.
In a binary system, a NS develops a quadrupole moment due to the tidal interaction with its companion~\cite{Flanagan:2007ix}.
The imprint of such tidal interactions was observed in the GW signal  GW170817~\cite{LIGOScientific:2017vwq} and led to constraints on 
the underlying NS equation of state (EOS)~\cite{LIGOScientific:2018hze,LIGOScientific:2018cki,Chatziioannou:2020pqz,Pradhan:2022rxs}.
These tidal interactions also give rise to oscillation modes of NS \cite{McDermott:1983ApJ...268..837M,McDermott:1988ApJ...325..725M,christensen1998lecture,Kokkotas:1999bd,Andersson:2021qdq}, in particular, the f-mode dynamical tides \cite{Will:1983vlw}, 
which have been argued to be important in inferring the NS EOS~\cite{Pratten:2021pro} in upcoming observing runs of present GW detectors. 
Neglecting the dynamical tidal effects can introduce significant biases in the estimation of the tidal deformability, consequently impacting the accuracy of the EOS inferences.
Moreover, the inclusion of dynamical tidal effects has been shown to improve the agreement between some GW models and numerical relativity simulations~\cite{Steinhoff:2016rfi,Hinderer:2016eia,Andersson:2019dwg,Schmidt:2019wrl} (see also Refs.~\cite{Akcay:2018yyh,Dietrich:2019kaq,Gamba:2022mgx,Gamba:2023mww}), and are necessary to fulfill the scientific objectives of the next generation ground-based GW observatories~\cite{Punturo:2010zza,Reitze:2019iox,Ackley:2020atn}. In this article, we aim to model the effects of such dynamic oscillations of NS on the dynamics of the binary system. 

We begin by considering the relativistic Lagrangian developed in Ref.~\cite{Steinhoff:2016rfi} for the quadrupole $f$-mode oscillation of a NS coupled to an external tidal field, describing the dynamical tidal (DT) effects,\footnote{
This representation of the Lagrangian is not unique, as it is subject to field redefinition: for instance, the last term could be equivalently replaced with a redefined $\lambda$ along with terms involving $(E_{\mu\nu})^2$ and $({\rm d}E_{\mu\nu}/{\rm d}\tau)^2$.
}
\begin{align}\label{eq_Lag_intro}
\mathcal{L}_{\rm DT}=\frac{z }{4\lambda\omega_f^2} \left[ \frac{c^2}{z^2}\frac{\dd Q_{\mu\nu}}{\dd\tau}\frac{\dd Q^{\mu\nu}}{\dd\tau} -\omega_f^2 Q_{\mu\nu}Q^{\mu\nu} \right] -\frac{z}{2} Q^{\mu\nu}  E_{\mu\nu}
- \kappa_{d} \frac{G_d^2 m^2}{c^6} \frac{z}{2}Q^{\mu\nu} \frac{\dd^2 E_{\mu\nu}}{\dd \tau^2}
\,,
\end{align}
where $\omega_f$ is the frequency of the mode, $\lambda$ is the tidal deformability \cite{Flanagan:2007ix}, $Q_{\mu\nu}$ is a symmetric trace-free tensor\footnote{We also impose a supplementary condition $Q_{\mu\nu}u^\mu=0$ on the quadrupole to project out the unphysical degrees of freedom.} that models the relativistic quadrupole moment of the star, $u^\mu=\dd x^\mu/\dd\tau$ is the four-velocity, $E_{\mu\nu}= - c^2 R_{\mu\alpha\nu\beta}u^\alpha u^\beta/z^2$ is the gravitoelectric field and $z=\sqrt{u^2}$ is the redshift factor. 
We work with the gravitational constant in $(d+1)$ spacetime dimensions written as $G_d = (\sqrt{4 \pi \exp(\gamma_{\rm E})} \, R )^{\epsilon} \, G_{N}$.
The last term in the above equation is the first non-minimal coupling that starts contributing to the conservative sector from 3PN order and $\kappa_{d}=(\sqrt{4 \pi \exp(\gamma_{\rm E})} \, R )^{-2\epsilon}\kappa$ is the {\it post-adiabatic Love number} in $(d+1)$ spacetime dimensions. 
We express $G_d$ and $\kappa_d$ in this particular form because later on we will employ the 
modified minimal subtraction scheme~\cite{Collins:1984xc}, and hence the appearance of the $4 \pi$, the Euler-Mascheroni constant $\gamma_{\rm E}$, and the (arbitrary) lenghtscale $R$.
In the adiabatic limit \cite{Hinderer:2007mb,Binnington:2009bb,Damour:2009vw}, where $\omega_f \rightarrow \infty$, the tides do not oscillate independently and are instead locked to the external tidal field as 
\begin{align}
    {Q}^{\mu\nu}=-\lambda {E}^{\mu\nu}- \lambda\kappa_d \frac{G_d^2 m^2}{c^6} \frac{\dd^2{E}^{\mu\nu}}{\dd \tau^2}\,.
\end{align}
The first term is the leading order term in the small frequency expansion of the conservative response function of the quadrupole~\cite{Chakrabarti:2013xza}, when sourced by external tidal field and the second term is the next-to-leading-order correction of it, known as the post-adiabatic term.
Substituting the above equation in Eq.~\eqref{eq_Lag_intro}, we obtain the Lagrangian for (post-)adiabatic tides (AT), which is given by,
\begin{align}\label{eq_Lag_ad}
\mathcal{L}_{\rm AT}=   \frac{z\lambda}{4}  E_{\mu\nu}E^{\mu\nu} 
+\lambda \kappa_d \frac{G_d^2 m^2}{c^6} \frac{z}{2} E_{\mu\nu} \frac{\dd^2 E^{\mu\nu}}{\dd \tau^2}
\,.
\end{align}
We can also write a similar adiabatic Lagrangian for the higher multipole moments which are studied in Ref.~\cite{Bini:2012gu}. See also Refs.~\cite{Thorne:1984mz,Zhang:1986cpa,Damour:1990pi,Binnington:2009bb,Damour:2009wj,Damour:2009vw,Henry:2019xhg}.
In general relativity, in addition to the relativistic gravitoelectric tides, we also get a new sector of gravitomagnetic tides \cite{Favata:2005da,Landry:2015cva,Pani:2018inf,Banihashemi:2018xfb,Poisson:2020eki,Poisson:2020mdi,Poisson:2020ify,Gupta:2020lnv} that are coupled to the odd-parity normal modes of the NS, modeled by the current-type multipole moments. For the adiabatic limit of the gravitomagnetic sector, see Ref.~\cite{Bini:2012gu}.
In a systematic EFT construction of dynamical tides, further couplings should be included, as outlined in~\cite{Gupta:2020lnv}, and applied to gravitomagnetic tides, which we leave for future work.

In this article, we quantify the influence of the dynamic tides on the behavior of a compact binary system. To achieve this goal, we employ effective field theory (EFT) techniques~\cite{Goldberger:2004jt} to analyze the inspiral phase of the binary, which occurs when its components are moving at nonrelativistic velocities and the orbital separation is large as compared to the length scale associated to each compact object.
In this context, we apply a perturbative approach that involves a series expansion based on powers of $v/c$, where $v$ represents the orbital velocity of the binary and $c$ is the speed of light. The virial theorem dictates that, the kinetic energy is ($-1/2$) times the potential energies of a bound state system. Thus, we set up the \emph{post-Newtonian} (PN) analysis, expanding in two perturbative parameters: $v/c$ and $G_N$, where $G_N$ denotes Newton's constant. Here, terms of the form $(v/c)^{n}$ are referred to as being of $(n/2)$PN order.
The PN analysis of the binary dynamics can be categorized into two sectors: the conservative sector, where emitted radiation is neglected and the orbital separation remains constant, and the radiative sector, where radiation carries away energy and momentum. At higher PN orders, these sectors can interact due to tail effects, originating from radiation being scattered by the orbital background curvature and affecting the orbital dynamics (see, e.g., Ref.~\cite{Blanchet:2013haa}.)
By employing the EFT approach, we can determine observables at any given PN order using diagrammatic methods, first proposed in~\cite{Goldberger:2004jt} and modern integration methods~\cite{Kol:2013ega,Foffa:2016rgu}, turning the problem into the determination of scattering amplitudes, which can be systematically obtained through the calculation of corresponding Feynman diagrams (see Refs.~\cite{Porto:2016pyg, Levi:2018nxp, Goldberger:2022ebt}, for reviews). 
Following the same computational strategy as developed in~\cite{Mandal:2022nty, Mandal:2022ufb, Mandal:2023lgy}, in order to compute the effective Lagrangian, we make use of an automated in-house code, interfaced to \texttt{QGRAF}~\cite{NOGUEIRA1993279}, for the diagram generation, to \texttt{xTensor}~\cite{xAct}, for tensor algebra manipulation, and to \texttt{LiteRed}~\cite{Lee:2013mka}, for the integral decomposition.

The effective Hamiltonian for conservative dynamical gravitoelectric tides was first computed in Refs.~\cite{Vines:2010ca,Steinhoff:2016rfi} up to 1PN and then was extended up to 2PN in Ref. \cite{Mandal:2023lgy}. The effects of spin and tides were analyzed together in Refs.~\cite{Gupta:2020lnv,Steinhoff:2021dsn,Gupta:2023oyy} 
for
gravitomagnetic tides. In the adiabatic limit, the 2PN effective Hamiltonian was computed in Refs.~\cite{Bini:2012gu,Henry:2019xhg} for both gravitoelectric and gravitomagnetic tides. 
Other works in PN theory can be found in Refs.~\cite{Vines:2011ud,Bini:2012gu,Vines:2010ca,Abdelsalhin:2018reg,Banihashemi:2018xfb,Landry:2018bil}.
In the post-Minkowskian (PM) expansion, where the perturbative series is controlled by $G_N$ alone, the adiabatic tidal corrections were studied to 3PM order 
in Refs.~\cite{Jakobsen:2022psy,Kalin:2020lmz}.
See also Refs.~\cite{Kalin:2020mvi,Bini:2020flp,Cheung:2020sdj,Haddad:2020que,Cheung:2020gbf,Bern:2020uwk,Mougiakakos:2022sic}.
Adiabatic tidal effects where also included in effective-one-body waveform models~\cite{Buonanno:1998gg,Buonanno:2000ef} in Refs.~\cite{Bini:2012gu,Damour:2009wj,Baiotti:2011am,Bernuzzi:2012ci,Bini:2014zxa} and in Refs.~\cite{Hinderer:2016eia,Steinhoff:2016rfi,Steinhoff:2021dsn} for the case of dynamical tides.
In this paper, we extend the state-of-the art of the \emph{analytic calculations of dynamical and adiabatic gravitoelectric tides in the conservative sector to 3PN order
}.

Within the diagrammatic EFT approach, higher-order perturbative corrections to the scattering amplitudes may contain divergent contributions. By adopting the dimensional regularization scheme, the divergent terms of multi-loop  Feynman integrals are parameterised by poles in $\epsilon=(d-3)$,
where $d$ is the continuous number of space-time dimension.
The so-called ultraviolet (UV) divergences 
can be absorbed in the redefinition of the free parameters of the theory, known as {\it renormalization}, thus yielding finite results. 
The process of renormalization can be understood as a coarse-graining transformation on the system \cite{PhysRevB.4.3174}.
One can understand the effects of renormalization in classical context using the Kadanoff's block-spin transformations \cite{Kadanoff:1966wm}.
See Ref.~\cite{Solodukhin:1997xn,Goldberger:2001tn} for renormalization in classical field theories coupled to external spatial sources. 

Besides the UV divergences, the computation of the conservative potential for a binary system may also involve infrared (IR) divergences. 
In the point particle sector, starting from the 4PN order, we encounter IR divergences in the conservative sector, which are due the artificial separation of conservative and dissipative dynamics~\cite{Foffa:2019yfl,Blumlein:2020pog}. However, in the computation of 3PN dynamical tides, we encounter only UV divergences. Part of these divergent terms can be removed by adding counterterms based on the world-line operators, which
eventually can be removed with field-redefinitions. This procedure is equivalent to finding a suitable canonical transformation (in the Hamiltonian picture) or adding a total derivative (in the Lagrangian picture)~\cite{Foffa:2019yfl}. We dub these divergences as spurious divergences as they do not have any effect on any physical observables. Additionally, we encounter UV divergences that are not spurious, which can be eliminated by the renormalization procedure, namely by introducing a new counterterm in the point particle Lagrangian, eventually yielding finite results. This counterterm affects the source coupling of the quadrupole moment with the gravitons.
Hence, the renormalization procedure introduces the running of the post-adiabatic Love number $\kappa$. 

Furthermore, we note that the same counterterm can also remove the UV divergences arising from the radiative sector \cite{Blanchet:1997jj,Goldberger:2009qd}. The appearance of an identical divergent term in both the conservative and radiative sector was observed in Ref.~\cite{Goldberger:2004jt} just for the case of spurious divergences, which can be removed through field redefinition completely.

Analogously, in the recent computation of 4PM observables due to scalar interactions \cite{Barack:2023oqp}, the renormalization of one of the scalar tidal Love numbers was necessary to obtain finite results. 
Renormalization and running for conservative and dissipative Love numbers was also seen in Ref. \cite{Saketh:2023bul} by determining the counter terms in the tidal response using BHPT (black-hole perturbation theory) for the case of Kerr black holes.
In this article, \emph{we present the renormalization of post-adiabatic Love number as a consequence of 
divergences appearing in the conservative sector of the gravitational two-body system.}

The paper is organized as follows. 
In Section~\ref{sec_setup}, we review
the description of tidally-interacting binaries in the EFT formalism.  
In Section~\ref{sec_Routine}, we present the algorithm used to compute the 3PN dynamic tidal potential.
In Section \ref{sec_renormalization}, we present the procedure of renormalization required for the post-adiabatic Love number to get a finite interaction Hamiltonian.
Our main result, the effective dynamical tidal Hamiltonian~\eqref{eq_ham_dy}, is presented in Section~\ref{sec_results}. 
In Section~\ref{sec_adiabatic}, we consider the adiabatic limit, and derive an effective adiabatic tidal Hamiltonian. 
We compute two gauge-independent observables: (i) the binding energy of a circular binary and (ii) the scattering angle for the hyperbolic encounter of two stars. 
Finally, we present our conclusions and avenues for future work in Section~\ref{sec_Conclusion}. 
This work is supplemented with three ancillary files: 
\texttt{Hamiltonian-DT.m}, containing the analytic expression of the Hamiltonian for the dynamic tides, \texttt{Hamiltonian-AT.m}, containing the analytic expression of the Hamiltonian for the adiabatic tides and \texttt{Poincare\_Algebra.m}, containing the result for the center of mass of the system which completes the Poincar\'e algebra and, hence, validates our results.

{\it Notation --}~The mostly negative signature for the metric is employed. Bold-face characters are used for three-dimensional variables, and normal-face font, for four-dimensional variables. The subscript ${(a)}$ labels the binary components on all the corresponding variables, like their position $\bm{x}_{(a)}$ and quadrupole moment $\bm{Q}_{(a)}$. An overdot indicates the time derivative, e.g., $\bm{v}_{(a)}=\dot{\bm{x}}_{(a)}$ is the velocity, $\bm{a}_{(a)}=\ddot{\bm{x}}_{(a)}$ the acceleration and $\dot{{\bm Q}}= \dd {\bm Q}/\dd t$. The separation between two objects is denoted by $\bm{r}=\bm{x}_{(1)}-\bm{x}_{(2)}$, with absolute value $r=|\bm{r}|$ and the unit vector along the separation is $\bm{n}=\bm{r}/r$.

\section{An EFT description of tides}
\label{sec_setup}

In this section, we review the formalism developed to model the dynamic tidal oscillations of compact objects in Ref.~\cite{Steinhoff:2016rfi}.
The Lagrangian given in equation \eqref{eq_Lag_intro} can be written as
\begin{align}
\mathcal{L}_{{\rm DT} (a)} &= c P_{(a)\mu\nu} \frac{\dd Q_{(a)}^{\mu\nu}}{\dd\tau}-
z_{(a)} \left[ \lambda_{(a)} \omega_{f(a)}^2 P_{(a)}^{\mu\nu}P_{(a)\mu\nu} + \frac{1}{4\lambda} Q_{(a)}^{\mu\nu}Q_{(a)\mu\nu} \right] \nonumber\\
&\quad -\frac{z_{(a)}}{2} Q_{(a)}^{\mu\nu}  E_{\mu\nu}- \kappa_{d(a)} \frac{G_d^2 m_{(a)}^2}{c^6} \frac{z_{(a)}}{2} Q_{(a)}^{\mu\nu}  \frac{d^2E_{\mu\nu}}{d\tau^2}   \,,
\label{eq_L_DT_intermediate}
\end{align}
where the conjugate momenta $P_{\mu\nu}$ with respect to the quadrupole moment are defined as,
\begin{align}
    P_{(a)\mu\nu}=\frac{1}{c}\frac{\partial \mathcal{L}}{\partial({\dd Q_{(a)}^{\mu\nu}}/{\dd \tau})}=\frac{c}{2\lambda \omega_f^2 z_{(a)}} \frac{\dd Q_{(a)\mu\nu}}{\dd\tau}\,.
\end{align}
The action for dynamical tides, written explicitly in terms of the physical
\footnote{The supplementary condition for the dynamical degrees of freedom in the rest frame of the star becomes:
$Q_{(a)}^{A0}=0\,, \quad \text{and}\quad P_{(a)}^{A0}=0$
where, we now explicitly see that $Q^{AB}_{(a)}$ and $P^{AB}_{(a)}$ are spatial tensors that encode only the physical degrees of freedom. Thus, we define the spatial tensor $Q_{(a)}^{AB}\delta_A^i\delta_B^j = \bm{Q}_{(a)}^{ij}$ and $P_{(a)}^{AB}\delta_A^i\delta_B^j = \bm{P}_{(a)}^{ij}$.
}
degrees of freedom, $\bm{Q}_{(a)}^{ij}$ and $\bm{P}_{(a)}^{ij}$, can be obtained by expressing the dynamical variables in the 
rest frame of each body
\footnote{
Different reference frames: 
(i) the general coordinate frame (denoted by Greek indices), 
(ii) the local Lorentz frame (denoted by small Latin indices), 
(iii) and the rest frame of the compact objects (denoted by capital Latin indices), and the Lorentz transformation, which boosts between the local Lorentz frame and the rest frame of the body is given by
\begin{align}
B^{a}_{(a)A}=\eta^{a}_{~A} + 2 \frac{u_{(a)}^a \delta_{A}^0}{z_{(a)}}-\frac{(u_{(a)}^a+z_{(a)}\delta^a_0)(u_{(a)A}+z_{(a)}\delta_A^0)}{z_{(a)}(z_{(a)}+u^a\delta_a^0)} \,. \nonumber
\end{align}
}.
This gives us the effective point-particle (``pp'') action
\begin{align}\label{eq_action_pp}
S_{\text{pp}} = \sum_{a=1,2} \int \frac{\dd \tau}{c} \left[ - m_{(a)} z_{(a)} c^2 + {\cal L}_{{\rm FD}(a)} + {\cal L}_{{\rm MQ}(a)} + {\cal L}_{\rm{EQ}(a)} +\mathcal{L}_{{\rm \ddot{E}Q}(a)} \right]\,.
\end{align}
The first term is simply the action for a point particle, while the remaining terms originate from the Lagrangian~\eqref{eq_L_DT_intermediate} as follows. 
The first term in Eq.~\eqref{eq_L_DT_intermediate} gives rise to,
\begin{align}
\mathcal{L}_{{\rm FD}(a)} &=
\bm{P}_{(a)}^{ij}\dot{\bm{Q}}_{(a)}^{ij}
+ c \left[ 
- u_{(a)}^\mu \omega_\mu^{ij} \left( \frac{\bm{S}_{Q(a)}^{ij}}{2} - \frac{\bm{S}_{Q(a)}^{ik} u^k_{(a)} u_{(a)}^j}{z_{(a)}(z_{(a)}+u_{(a)}^a \delta_a^0)} \right)
- u_{(a)}^\mu \omega_\mu^{ai}\delta_a^0 \bm{S}_{Q(a)}^{ij} \frac{u^j_{(a)}}{z_{(a)}}
\right.
\nonumber \\
&\qquad\qquad\qquad\quad\,\,
\left. 
+ \frac{\bm{S}_{Q(a)}^{ij}u^i_{(a)}}{z_{(a)}(z_{(a)}+u_{(a)}^a \delta_a^0)}\frac{\dd u^j_{(a)}}{\dd \tau}
\right] \,,
\end{align}
which describes frame-dragging (``FD'') effects on the quadrupole moment of each binary component, where, the ``tidal spin'' tensor
$\bm{S}_{Q(a)}^{ij} = 2 \, (\bm{Q}_{(a)}^{ki}\bm{P}_{(a)}^{jk}-\bm{Q}_{(a)}^{kj}\bm{P}_{(a)}^{ik})$ \,,
which describes the angular momentum of the dynamical quadrupole moment.
The second term in Eq.~\eqref{eq_L_DT_intermediate} yields,
\begin{align}
\mathcal{L}_{{\rm MQ}(a)} = -z_{(a)} \left[ \lambda_{(a)} \omega_{f(a)}^2 \bm{P}_{(a)}^{ij}\bm{P}_{(a)}^{ij} + \frac{1}{4\lambda_{(a)}} \bm{Q}_{(a)}^{ij}\bm{Q}_{(a)}^{ij} \right]
= -z_{(a)} \bm{M}_{Q(a)} \,.
\label{eq_L_MQ}
\end{align}
This term governs the dynamics of the quadrupole moment, which, by the second equality, can be described as a time-dependent effective mass term for quadrupole moment (``MQ'').
Finally, the last two terms in Eq.~\eqref{eq_L_DT_intermediate} results in,
\begin{align}\label{eq_L_source}
\mathcal{L}_{{\rm EQ}(a)} =
- \frac{z_{(a)}}{2}\bm{Q}_{(a)}^{ij} \bm{E}_{(a)}^{ij} \, \quad \quad \text{and} \quad \quad \mathcal{L}_{{\rm \ddot{E}Q}(a)} =
- \kappa_{d(a)} \frac{G_d^2 m_{(a)}^2}{c^6} \frac{z_{(a)}}{2} \bm{Q}_{(a)}^{ij}\ddot{\bm{E}}_{(a)}^{ij} \,.
\end{align}
These terms act as a driving source for the quadrupole moment's dynamics and are induced on each of the binary components by the gravitoelectric tidal field $\bm{E}_{(a)}^{ij}=B^a_{(a)i}B^b_{(a)j}  e^\mu_{~a} e^\nu_{~b} E_{\mu\nu}$ of its companion.

Here the Poisson bracket of dynamical quantities i.e. the dynamic quadrupole $\textbf{Q}^{ij}$ and its conjugate momenta $\bm{P}^{ij}$ is given as
\begin{align}
\label{eq_Q_Poisson_bracket}
    \{ \bm{Q}^{ij}, \bm{P}^{kl} \}&= \frac{1}{2} \left(\delta^{ik}\delta^{jl}+\delta^{il}\delta^{jk}\right)-\frac{1}{3}\delta^{ij}\delta^{kl} \,, 
\end{align}
which then implies a SO(3) angular momentum algebra for the tidal spin tensor $\bm{S}_Q^{ij}$.

\section{Computational algorithm}
\label{sec_Routine}

\def\IREFT{IR-EFT }
\def\UVEFT{UV-EFT }

\subsection{The effective Lagrangian}

In this section, we present the used computational algorithm. The resulting potential will then be used in the next section to obtain the effective two-body Hamiltonian.
The dynamics of the gravitational field $g_{\mu\nu}$ is given by the Einstein-Hilbert action along with a harmonic gauge fixing term in $d+1$ spacetime dimensions,
\begin{align}\label{eq_action_EH}
S_{\text{EH}} = -\frac{c^4}{16 \pi G_d} \int \dd^{d+1} x~\sqrt{g} ~\mathcal{R} + \frac{c^4}{32 \pi G_d } \int \dd^{d+1} x~\sqrt{g} ~g_{\mu\nu}~\Gamma^\mu \Gamma^\nu \, ,
\end{align}
where $\Gamma^\mu=\Gamma^{\mu}_{\rho\sigma}g^{\rho\sigma}$, 
$\Gamma^{\mu}_{\rho\sigma}$ is the Christoffel symbol, $\mathcal{R}$ is the Ricci scalar, and $g$ is the metric determinant.

For the conservative dynamics of the system, we decompose the metric as $g_{\mu\nu}=\eta_{\mu\nu}+H_{\mu\nu}$, where $H_{\mu\nu}$ is the potential graviton.
We use the Kaluza-Klein parametrization to decompose the metric, where the 10 degrees of freedom of $H_{\mu\nu}$ are encoded in three fields: a scalar $\bm{\phi}$, a three-dimensional vector $\bm{A}$ and a three-dimensional symmetric rank two tensor $\bm{\sigma}$~\cite{Kol:2007bc,Kol:2007rx}. In this parametrization, we write the metric as
\begin{equation}
g_{\mu\nu} = 
\begin{pmatrix}
e^{2\bm{\phi}/c^2} \,\,\, & -e^{2\bm{\phi}/c^2} {\bm{A}_j}/{c^2}\\
-e^{2\bm{\phi}/c^2} {\bm{A}_i}/{c^2} \,\,\,\,\,\, & -e^{-2\bm{\phi}/((d-2)c^2)}\bm{\gamma}_{ij}+e^{2\bm{\phi}/c^2} {\bm{A}_i}{\bm{A}_j}/{c^4}  
\end{pmatrix}\,,
\quad {\rm with} \quad
\bm{\gamma}_{ij}=\bm{\delta}_{ij}+\bm{\sigma}_{ij}/c^2 \,.
\end{equation}

We can now obtain the effective action for the binary by integrating out the gravitational degrees of freedom as follows,
\begin{align}
\exp \left[{\ii \, \int \dd t ~ \mathcal{L}_{\text{eff}}}\right] = \int \DD \bm{\phi} \, \DD \bm{A}_i \, \DD \bm{\sigma}_{ij}\, \exp[\ii \, (S_{\text{EH}}+S_{\text{pp}})] \, ,
\end{align}
where the Einstein-Hilbert action is given by Eq.~\eqref{eq_action_EH} and the point-particle action is given by Eq.~\eqref{eq_action_pp}.
To perform this integration, we first decompose the effective Lagrangian ${\cal L}_{\rm eff}$ as 
\begin{equation}
\mathcal{L}_{\rm eff} =  \mathcal{K}_{\rm eff} - \mathcal{V}_{\rm eff} \,,
\end{equation}
where $\mathcal{K}_{\rm eff}$ is an effective kinetic term, which does not dependent on any integration of potential graviton (i.e., it is independent of integration over ${\bm \phi}$, ${\bm A}$, and ${\bm \sigma}$). We can compute $\mathcal{K}_{\rm eff}$ directly up to the required PN order. Explicitly, we decompose $\mathcal{K}_{\rm eff}$ in a point-particle, a frame-dragging, and a ``quadrupole mass'' contribution, i.e., ${\cal K}_{\rm eff} = {\cal K}_{\rm pp} + {\cal K}_{\rm FD} + {\cal K}_{\rm MQ}$,
\begin{subequations}
\label{eq_K_all}
\begin{align}\label{eq_Kpp}
\mathcal{K}_{\rm pp} =& \sum_{a=1,2} m_{(a)} 
\left[
\frac{1}{2}  \bm{v}_a^2 + \frac{1}{8}  \bm{v}_{(a)}^4 \left(\frac{1}{c^2}\right)  + \frac{1}{16}  \bm{v}_{(a)}^6 \left(\frac{1}{c^4}\right) + \frac{5}{128}  \bm{v}_{(a)}^8 \left(\frac{1}{c^6}\right) 
\right]
+ \mathcal{O}\left(\frac{1}{c^8}\right)\,, \\
\label{eq_KSO}
\mathcal{K}_{\rm FD} =& \sum_{a=1,2} 
\left\{ \bm{P}_{(a)}^{ij}\dot{\bm{Q}}_{(a)}^{ij} + \bm{S}_{Q(a)}^{ij}\bm{v}_{(a)}^i\bm{a}_{(a)}^j 
\left[
\frac{1}{2} \left(\frac{1}{c^2}\right)+ \frac{3}{8}  \bm{v}_{(a)}^2 \left(\frac{1}{c^4}\right)  + \frac{5}{16}  \bm{v}_{(a)}^4 \left(\frac{1}{c^4}\right)
\right] \right\} + \mathcal{O}\left(\frac{1}{c^{8}}\right) \,, \\
\label{eq_KMQ}
\mathcal{K}_{\rm MQ} =& \sum_{a=1,2} \bm{M}_{(a)} 
\left[ 
1 +\frac{1}{2}  \bm{v}_a^2 \left(\frac{1}{c^2}\right)  + \frac{1}{8}  \bm{v}_{(a)}^4 \left(\frac{1}{c^4}\right)  + \frac{1}{16}  \bm{v}_{(a)}^6 \left(\frac{1}{c^6}\right) \right]
+ \mathcal{O}\left(\frac{1}{c^8}\right)\,.
\end{align}
\end{subequations}

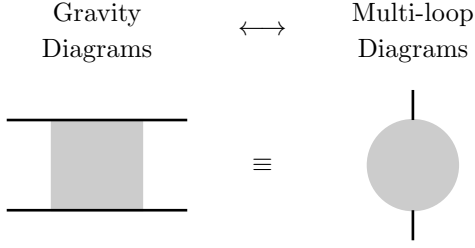
\begin{figure}[H]
	\centering
	\begin{tikzpicture}[line width=1 pt, scale=0.4]
	\begin{scope}[shift={(-7,0)}]
	\filldraw[color=gray!40, fill=gray!40, thick](0,0) rectangle (3,3);
	\draw (-1.5,0)--(4.5,0);
	\draw (-1.5,3)--(4.5,3);
	\node at (1.5,6.5) {Gravity};	
	\node at (1.5,5.3) {Diagrams};	
	\end{scope}
	\begin{scope}[shift={(0,0)}]
	\node at (0,6) {$\longleftrightarrow$};
	\node at (0,1.5) {$\equiv$};
	\end{scope}
	\begin{scope}[shift={(5,0)}]
	\filldraw[color=gray!40, fill=gray!40, thick](0,1.5) circle (1.5);
	\draw (0,3)--(0,4);
	\draw (0,0)--(0,-1);	
	\node at (0,6.5) {Multi-loop};				
	\node at (0,5.3) {Diagrams};				
	\end{scope}		
	\end{tikzpicture}
	\caption{The diagrammatic correspondence between the four-point EFT-Gravity graphs
and the two-point quantum-field-theory (QFT) graphs.}
	\label{fig_relation_bet_diags}
\end{figure}

The terms that are obtained after performing the explicit integral are collectively denoted by the potential $\mathcal{V}_{\rm eff}$. These terms are computed by summing over the connected Feynman diagrams without graviton loops, as shown below,
\begin{align}
\mathcal{V}_{\text{eff}} = \ii \,
\lim_{d\rightarrow 3} 
\int \frac{\dd^d \bm{p}}{(2\pi)^d}~
e^{\ii \, \bm{p}\cdot (\bm{x}_{(1)}-\bm{x}_{(2)})}
\, 		
\parbox{25mm}{
	\begin{tikzpicture}[line width=1 pt,node distance=0.4 cm and 0.4 cm]
	\coordinate[label=left: ] (v1);
	\coordinate[right = of v1] (v2);
	\coordinate[right = of v2] (v3);
	\coordinate[right = of v3] (v4);
	\coordinate[right = of v4, label=right: \tiny$(2)$] (v5);
	\coordinate[below = of v1] (v6);
	\coordinate[below = of v6, label=left: ] (v7);
	\coordinate[right = of v7] (v8);
	\coordinate[right = of v8] (v9);
	\coordinate[right = of v9] (v10);
	\coordinate[right = of v10, label=right: \tiny$(1)$] (v11);
	\fill[black!25!white] (v8) rectangle (v4);
	\draw[fermionnoarrow] (v1) -- (v5);
	\draw[fermionnoarrow] (v7) -- (v11);
	\end{tikzpicture}
}
\, ,
\label{eq:effective_lagrangian}
\end{align}
where $\bm{p}$ is the linear momentum transferred between the two bodies. 
For this, we begin with generating all the topologies that correspond to graviton exchanges between the worldlines of the two compact objects. 
There is 1 topology at tree-level ($G_N$), 2 topologies at one-loop ($G_N^2$), 9 topologies at two-loop ($G_N^3$), and 32 topologies at three-loop ($G_N^4$).
We then dress these topologies with the Kaluza-Klein fields $\bm{\phi}$, $\bm{A}$ and $\bm{\sigma}$. The number of diagrams\footnote{The diagrams which can be
obtained from the change in the label $1\leftrightarrow 2$, are not counted as separate diagrams.} appearing in the point-particle sector is given in Table~\ref{tbl_no_diag_non_spinning}, whereas that in the tidal sector are given in Tables~\ref{tbl_no_diag_tidalEQ},~\ref{tbl_no_diag_tidalFD}, ~\ref{tbl_no_diag_tidalMQ} and \ref{tbl_no_diag_tidalpAd}.
We use an in-house code that uses tools from \texttt{EFTofPNG}~\cite{Levi:2017kzq} and \texttt{xTensor}~\cite{xAct}, for the tensor algebra manipulation, and \texttt{LiteRED} \cite{Lee:2013mka}, for the integration-by-parts reduction, to compute these Feynman diagrams. 
This reduction recasts the Feynman diagrams in terms of two point massless master integrals~\cite{Foffa:2016rgu} as shown in Fig.~\ref{fig_relation_bet_diags}.
Once the exact expressions for the master integrals are substituted, we perform a Fourier transform to obtain the position-space effective potential ${\cal V}_{\rm eff}$. The details of the algorithm and the expressions for the master integrals up to three loops can be found in Ref.~\cite{Mandal:2022nty}.

After carrying out all these steps, the effective potential can be decomposed into
a point-particle and a dynamical tide contribution, i.e., ${\cal V}_{\rm eff} = {\cal V}_{\rm pp} + {\cal V}_{\rm DT}$, where
\begin{subequations}
\begin{align}
\mathcal{V}_{\rm pp} &= \mathcal{V}_{\rm N} + \left(\frac{1}{c^2}\right) \mathcal{V}_{\rm 1PN} + \left(\frac{1}{c^4}\right) \mathcal{V}_{\rm 2PN} + \left(\frac{1}{c^6}\right) \mathcal{V}_{\rm 3PN} + \mathcal{O}\left(\frac{1}{c^8}\right)\, , \\
\mathcal{V}_{\rm DT} &= \sum_{n = 0}^{3} \left(\frac{1}{c^2}\right)^n 
\left(
{\cal V}^{\rm EQ}_{n{\rm PN}}
+ {\cal V}^{\rm FD}_{n{\rm PN}}
+ {\cal V}^{\rm MQ}_{n{\rm PN}}
\right) 
+ \left(\frac{1}{c^6}\right) \mathcal{V}^{\rm \ddot{E}Q}_{\rm 3PN} + \mathcal{O}\left(\frac{1}{c^6}\right) \,,
\end{align}
\end{subequations}
and we remark that ${\cal V}_{\rm DT}$ has contributions due to the driving source, the ``quadrupole-mass'' and the frame-dragging terms, and the post adiabatic terms.

\begin{table}
	\begin{subtable}[H]{0.49\textwidth}
    \centering
		\begin{tabular}{|c|c|c|c|}
			\hline
			Order                & Diagrams            & Loops       & Diagrams  \\ \hline
			\multirow{4}{*}{3PN} & \multirow{4}{*}{130} & 0    & 9 \\ \cline{3-4} 
			&                     & 1 & 38 \\ \cline{3-4} 
			&                     & 2  & 75 \\ \cline{3-4}
            &                     & 3  & 8 \\ \hline
		\end{tabular}
		\caption{Point particle sector}
		\label{tbl_no_diag_non_spinning}
	\end{subtable}
	\begin{subtable}[H]{0.49\textwidth}
    \centering
		\begin{tabular}{|c|c|c|c|}
			\hline
			Order                & Diagrams            & Loops      & Diagrams  \\ \hline
			\multirow{4}{*}{3PN} & \multirow{4}{*}{168} & 0    &  9\\ \cline{3-4} 
			&                     & 1 &  43\\ \cline{3-4} 
			&                     & 2  &  101\\ \cline{3-4}
            &                     & 3  &  15\\ \hline
		\end{tabular}
		\caption{EQ sector}
		\label{tbl_no_diag_tidalEQ}
	\end{subtable}
	
	\vspace{0.5cm}
	\begin{subtable}[H]{0.49\textwidth}
    \centering
		\begin{tabular}{|c|c|c|c|}
			\hline
			Order                & Diagrams            & Loops      & Diagrams  \\ \hline
			\multirow{3}{*}{3PN} & \multirow{3}{*}{100} & 0   & 8 \\ \cline{3-4} 
			&                     & 1 &  36 \\ \cline{3-4} 
			&                     & 2  & 56 \\ \hline
		\end{tabular}
		\caption{FD sector}
		\label{tbl_no_diag_tidalFD}
	\end{subtable}
    \begin{subtable}[H]{0.49\textwidth}
    \centering
		\begin{tabular}{|c|c|c|c|}
			\hline
			Order                & Diagrams            & Loops      & Diagrams  \\ \hline
			\multirow{3}{*}{3PN} & \multirow{3}{*}{21} & 0   & 6 \\ \cline{3-4} 
			&                     & 1 & 10 \\ \cline{3-4} 
			&                     & 2  & 5 \\ \hline
		\end{tabular}
		\caption{MQ sector}
		\label{tbl_no_diag_tidalMQ}
	\end{subtable}

    \vspace{0.5cm}
    \begin{subtable}[H]{1\textwidth}
    \centering
		\begin{tabular}{|c|c|c|c|}
			\hline
			Order                & Diagrams            & Loops      & Diagrams  \\ \hline
			3PN & 1 & 0   &  1 \\ \hline
		\end{tabular}
		\caption{${\rm \ddot{E}Q}$ sector}
		\label{tbl_no_diag_tidalpAd}
	\end{subtable}
	\label{tbl_no_diag}
	\caption{Number of Feynman diagrams contributing different sectors.
 }
\end{table}

\subsection{Removal of higher order time derivatives}

The potential ${\cal V}_{\rm eff}$ computed in the previous section, is a function of the dynamical variables $\bm{x}_{(a)}$, $\bm{Q}_{(a)}$, and $\bm{S}_{Q(a)}$ and $\bm{M}_{Q(a)}$, and their higher order time derivatives.
The first and higher-order time derivatives of $\bm{Q}_{(a)}$, $\bm{S}_{Q(a)}$, and $\bm{M}_{Q(a)}$ are removed using integration by parts, while second and higher-order time derivatives of $\bm{x}_{(a)}$ 
are removed using a coordinate transformation $\bm{x}_{(a)} \rightarrow \bm{x}_{(a)} + \delta \bm{x}_{(a)}$ \cite{Levi:2014sba,Schafer:1984mr,Damour:1990jh,Damour:1985mt,Barker:1980spx}. 
This coordinate transformation changes the Lagrangian as
\begin{align}
\delta \mathcal{L} = 
\frac{\delta \mathcal{L}}{\delta \bm{x}_{(a)}^i} \, \delta \bm{x}_{(a)}^i + \mathcal{O} (\delta \bm{x}_{(a)}^2) \,,
\end{align}
where $\delta \bm{x}_{(a)}$ is chosen such that it removes the undesirable terms from our final Lagrangian.
In our case, the process of removing the higher order time derivatives using a coordinate transformation is equivalent to the substitution of the equation of motion for the acceleration $\bm{a}_{(a)}$ and its higher order time derivatives back into the Lagrangian \cite{Damour:1990jh}.
This procedure converts the Lagrangian $\mathcal{L}_{\rm eff}$ into the Lagrangian $\bar{\mathcal{L}}_{\rm eff}$ which depends only on $\bm{x}_{(a)}$, $\bm{v}_{(a)}$, and $\bm{Q}_{(a)}$, but still contains divergent terms.

\subsection{Removal of spurious divergences}
After the removal of the higher order time derivatives, the effective Lagrangian $\bar{\mathcal{L}}_{\rm eff}$ contains several divergent terms. These divergent terms are generated by the two topologies as shown in Fig.~\ref{fig_div_diag}.
Part of these divergent terms can be removed by adding counter-terms based on the world-line operators, which ultimately can be removed by using field-redefinitions. So, this procedure becomes equivalent to adding total derivative terms to the Lagrangian or finding a canonical transformation on the corresponding Hamiltonian. As a result these divergences do not have any effect on the physical observable and are dubbed as spurious divergences.\footnote{The spurious divergences appearing in 3PN order can also be removed from the equation of motion \cite{Blanchet:2003gy}.}
For the removal of the 3PN spurious divergence in the point particle sector, 
following \cite{Foffa:2011ub}, we can add a total derivative term 
\begin{align}
\mathcal{L}_{\text{TD1}}=\left(\frac{1}{c^6}\right)\frac{{\rm d}}{{\rm d}t}\left[\frac{G_N^3}{r^2}\Big(c_1 \left(\bm{v}_{(1)} \cdot \bm{n}\right) + c_2 \left(\bm{v}_{(2)} \cdot \bm{n}\right)\Big)\right]\, ,
\end{align}
and gaining insights from \cite{Levi:2014sba}, we choose another total derivative term
\begin{align}
\mathcal{L}_{\text{TD2}}&=\left(\frac{1}{c^6}\right)\frac{{\rm d}}{{\rm d}t}\Bigg\{ \frac{G_N^3}{r^4}\left[c_3   \Big(\bm{Q}_{(1)}^{ij} \bm{v}_{(1)}^i \bm{n}^j\Big) + c_4 \Big(\bm{Q}_{(1)}^{ij} \bm{v}_{(2)}^i \bm{n}^j \Big) + c_5 \Big( \bm{Q}_{(2)}^{ij} \bm{v}_{(1)}^i \bm{n}^j \Big) + c_6 \Big(\bm{Q}_{(2)}^{ij} \bm{v}_{(2)}^i \bm{n}^j\Big) \right]\nonumber\\
& +\frac{G_N^3}{r^4}\Bigg[ \Big(\bm{Q}_{(1)}^{ij} \bm{n}^i \bm{n}^j\Big) \Big(c_7 \left(\bm{v}_{(1)} \cdot \bm{n}\right) + c_8  \left(\bm{v}_{(2)} \cdot \bm{n}\right)\Big) + \Big(\bm{Q}_{(2)}^{ij} \bm{n}^i \bm{n}^j\Big) \Big(c_9 \left(\bm{v}_{(1)} \cdot \bm{n}\right) + c_{10} \left(\bm{v}_{(2)} \cdot \bm{n}\right)\Big) \Bigg] \Bigg\}\, ,
\end{align}
to remove the spurious divergence in the 3PN tidal sector.
Here the coefficients are represented in the notation,
\begin{align}
c_n \equiv c_{\epsilon n} \frac{1}{\epsilon} + c_{Ln} \log\left(\frac{r}{R}\right)\, , \, \, \text{for} \,\, n=1,2,\cdots 10 \, ,
\end{align}
and their particular values are given as,
\begin{align}
c_{\epsilon 1}&= \frac{1}{3} \left(4  m_{(1)}^3 m_{(2)}- m_{(1)} m_{(2)}^3\right), & c_{\epsilon 2}&= \frac{1}{3} \left( m_{(1)}^3 m_{(2)}-4  m_{(1)} m_{(2)}^3\right)\, ,\nonumber\\
c_{\epsilon 3}&= -\frac{107}{35}   m_{(1)}^2 m_{(2)}^2 , & c_{\epsilon 4}&= \frac{107}{35}   m_{(1)}^3 m_{(2)} \, ,\nonumber\\
c_{\epsilon 5}&= -\frac{107}{35}   m_{(1)} m_{(2)}^3 , & c_{\epsilon 6}&= \frac{107}{35}   m_{(1)}^2 m_{(2)}^2 \, ,\nonumber\\
c_{\epsilon 7}&= \frac{107}{14}   m_{(1)}^2 m_{(2)}^2 , & c_{\epsilon 8}&= -\frac{107}{14}   m_{(1)}^3 m_{(2)} \, ,\nonumber\\
c_{\epsilon 9}&= \frac{107}{14}   m_{(1)} m_{(2)}^3 , & c_{\epsilon 10}&= -\frac{107}{14}   m_{(1)}^2 m_{(2)}^2 \, ,
\end{align}
and
\begin{align}
    c_{Li}&=-3 \,c_{\epsilon i} ~~~ \text{ for }i=1,2 \, 
    &c_{Li}&=-2 \, c_{\epsilon i} ~~~ \text{ for }i=3,\dots,10 \,.
\end{align}
Now the modified Lagrangian $\bar{\mathcal{L}}_{\text{eff}} + \mathcal{L}_{\text{TD1}} + \mathcal{L}_{\text{TD2}}$ is free of all the spurious divergences and still contains divergent terms in the 3PN tidal sector, which requires renormalization and we analyse it in the next section.

\section{Renormalization}
\label{sec_renormalization}

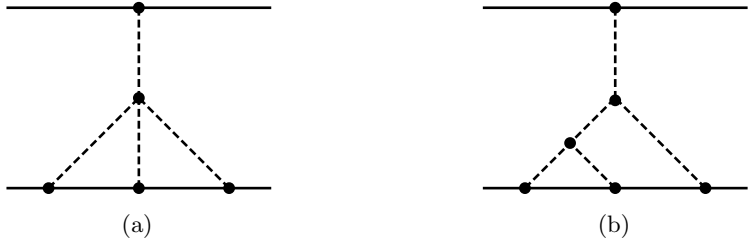
\begin{figure}[t]
	\centering
	\begin{subfigure}{0.4\textwidth}
		\centering
        	\begin{tikzpicture}[line width=1 pt, scale=0.8]
        	\begin{scope}[shift={(-7,0)}]
        	\draw (-2.2,0)--(2.2,0);
        	\draw (-2.2,3)--(2.2,3);
         	\draw[bgrav] (-1.5,0)--(0,1.5);
            \draw[bgrav] (1.5,0)--(0,1.5);
            \draw[bgrav] (0,0)--(0,1.5);
            \draw[bgrav] (0,3)--(0,1.5);
            \draw[black,fill=black] (0,0) circle (.5ex);
        	\draw[black,fill=black] (-1.5,0) circle (.5ex);
        	\draw[black,fill=black] (1.5,0) circle (.5ex);
        	\draw[black,fill=black] (0,3) circle (.5ex);
            \draw[black,fill=black] (0,1.5) circle (.5ex);
        	\end{scope}
        	\end{tikzpicture}
		\caption{}
	\end{subfigure}
	\begin{subfigure}{0.4\textwidth}
		\centering
        	\begin{tikzpicture}[line width=1 pt, scale=0.8]
        	\begin{scope}[shift={(-7,0)}]
        	\draw (-2.2,0)--(2.2,0);
        	\draw (-2.2,3)--(2.2,3);
         	\draw[bgrav] (-1.5,0)--(0,1.5);
            \draw[bgrav] (1.5,0)--(0,1.5);
            \draw[bgrav] (0,0)--(-0.75,0.75);
            \draw[bgrav] (0,3)--(0,1.5);
            \draw[black,fill=black] (0,0) circle (.5ex);
        	\draw[black,fill=black] (-1.5,0) circle (.5ex);
        	\draw[black,fill=black] (1.5,0) circle (.5ex);
        	\draw[black,fill=black] (0,3) circle (.5ex);
            \draw[black,fill=black] (0,1.46) circle (.5ex);
            \draw[black,fill=black] (-0.75,0.75) circle (.5ex);
        	\end{scope}
        	\end{tikzpicture}
		\caption{}
	\end{subfigure}
    \caption{The topologies that give divergent contribution in the tidal sector at 3PN. Solid lines represent the compact object's worldlines and the dashed lines represent the potential gravitons $H_{\mu\nu}$. 
    }
	\label{fig_div_diag}
\end{figure}

In this section, we first analyze the divergent terms present in the effective Lagrangian, which cannot be eliminated through the addition of total derivatives. These divergent terms are generated by the two topologies shown in Fig.~\ref{fig_div_diag}. We show that to eliminate these remaining divergences, a renormalization procedure is necessary, where the divergent contributions are absorbed into the bare post-adiabatic Love number $\kappa^{\rm B}$. As a consequence, we obtain a renormalized post-adiabatic Love number $\kappa (R)$, which exhibits a nontrivial renormalization group flow.

\subsection{Analysis of divergent terms}

First, we conduct a complete validity check of our computation, following the observation in Ref.~\cite{Steinhoff:2016rfi}. Specifically, the Hamiltonians $\widetilde{\mathcal{H}}^{\text{EQ}}$ and $\widetilde{\mathcal{H}}^{\text{FD}}$ exhibit similarities to the spin-induced quadrupole ($\widetilde{\mathcal{H}}^{\text{ES}^2}$) and spin-orbit Hamiltonians ($\widetilde{\mathcal{H}}^{\text{SO}}$) when certain replacements are applied. 
Our analysis starts with $\bar{\mathcal{L}}_{\text{eff}} + \mathcal{L}_{\text{TD1}}$, from which we derive the corresponding Hamiltonians $\widetilde{\mathcal{H}}^{\text{EQ}}_{3{\rm PN}}$ and $\widetilde{\mathcal{H}}^{\text{FD}}_{3{\rm PN}}$. These Hamiltonians are free of divergent and logarithmic terms in the point particle sector but contain such terms in the tidal sectors, respectively. On the other hand, following~\cite{Mandal:2022nty, Mandal:2022ufb}, we obtain the $\widetilde{\mathcal{H}}^{\text{SO}}_{\text{\NNLO}}$ and $\widetilde{\mathcal{H}}^{\text{ES}^2}_{\text{\NNNLO}}$, which are free of divergent and logarithmic terms in the point particle sector, but contains such terms in the spinning sector.
Further investigations led us to find that these Hamiltonians are equivalent to each other up to a canonical transformation, with $\widetilde{\mathcal{H}}^{\text{EQ}}_{3{\rm PN}}$ being equivalent to $\widetilde{\mathcal{H}}^{\text{ES}^2}_{\text{\NNNLO}}$, and $\widetilde{\mathcal{H}}^{\text{FD}}_{3{\rm PN}}$ being equivalent to $\widetilde{\mathcal{H}}^{\text{SO}}_{\text{\NNLO}},$ provided certain replacements are applied.
This result serves as a robust consistency check, confirming the accuracy of our computation of the effective Lagrangian and the procedure of removing the higher-order time derivatives.
However, an important distinction arises, while eliminating the divergent terms using a canonical transformation, between the tidal and the spinning sector.
Notably, the kinetic term of $\bm{Q}^{ij}$ begins at 0PN order, whereas the kinetic term of $\bm{S}^{ij}$ starts at 0.5PN order. This discrepancy has a crucial consequence while computing the canonical transformation. The contributions from the Poisson bracket of the spin tensor at \NNNLO can be ignored (see Ref. \cite{Mandal:2022ufb}, Sec.~4.3.2), whereas the equivalent contributions from the Poisson bracket for the quadrupole tensor (Eq.~\eqref{eq_Q_Poisson_bracket}) at 3PN order cannot be ignored. This distinction accounts for the origin of the residual divergent terms present in the tidal sector while all the divergent pieces in the spinning sector can be removed by finding a suitable canonical transformation.

\begin{figure}[t]
		\centering
        	\begin{tikzpicture}[line width=1 pt, scale=0.8]
        	\begin{scope}[shift={(-7,0)}]
        	\draw (-1.5,0)--(1.5,0);
        	\draw (-1.5,3)--(1.5,3);
         	\draw[bgrav] (0,0)--(0,3);
            \draw[black,fill=black] (0,0) circle (.5ex);
        	
            \path (0,3) pic[black] {cross=5pt};
            \draw (0,3) circle (1.5ex);
        	\end{scope}
        	\end{tikzpicture}
    \caption{Counter-term diagram. The the cross symbol stands for the insertion of the counterterm constant $\delta_{\kappa (a)}$. 
    Solid lines represent the compact object's worldlines and the dashed lines represent the potential gravitons $H_{\mu\nu}$.}
	\label{fig_CT}
\end{figure}
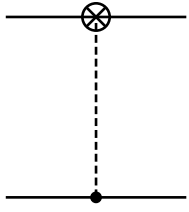

After the removal of all the spurious divergent terms, we obtain the following residual divergent term 
\begin{align}\label{eq_div_terms}
    \Big(\bar{\mathcal{L}}_{\text{eff}} + \mathcal{L}_{\text{TD1}} + \mathcal{L}_{\text{TD2}} \Big)_{1/\epsilon}= \Bigg( \frac{107}{105} m_{(1)}^2 G_N^2\frac{1}{c^6} \frac{1}{\epsilon}\Bigg) \Bigg( \frac32\frac{G_N m_{(2)}}{r^3}\Big(\ddot{\bm{Q}}_{(1)}^{ij}  \bm{n}^i  \bm{n}^j \Big) \Bigg) + ( 1 \leftrightarrow 2 ) \,,
\end{align}
where we perform some algebraic manipulations to present the expression in a compact form. 
We observe that the divergent terms in Eq.~\eqref{eq_div_terms} cannot be expressed as a total derivative and the tail effects in the case of dynamic tides start at 4PN order. 
So, we absorb the divergence in one of the coupling constants of the Lagrangian given in Eq. \eqref{eq_Lag_intro}, by designing a counterterm, which will eventually lead to finite results.
By examining the structure of terms in Eq.~\eqref{eq_div_terms}, 
we identify the following counterterm, which we can add to the point particle action \eqref{eq_action_pp}.
\begin{align}\label{eq_Lag_CT}
{\cal L}_{\rm{CT}(a)}=- \delta_{\kappa(a)} \frac{G_N^2 m_{(a)}^2}{c^6} \frac{z_{(a)}}{2}\bm{Q}_{(a)}^{ij}\ddot{\bm{E}}_{(a)}^{ij} \,,
\end{align}
where $\delta_{\kappa(a)}$ contains the divergence. 
The counter term Lagrangian contributes through a tree level diagram  at this order as shown in Fig.~\ref{fig_CT}, which renders the result finite, eventually.
Now, the particular value of $\delta_{\kappa(a)}$ that cancels the divergent terms presented in the effective Lagrangian \eqref{eq_div_terms} is,
\begin{align}
        \delta_{\kappa(a)} = -\frac{107}{105}  \frac{1}{\epsilon}\,.
\end{align}
We note that the counterterm has the structure similar to ${\cal L}_{{\rm \ddot{E}Q}(a)}$, while the divergence produced in the Eq. \eqref{eq_div_terms} is sourced by the ${\cal L}_{{\rm EQ}(a)}$.
The addition of the counterterm to the point particle action in Eq. \eqref{eq_action_pp}, leads to the renormalization of the post-adiabatic Love number
\begin{align}
\label{eq_kappa_delta}
    \kappa_{d(a)} = \Big(\sqrt{4 \pi \exp(\gamma_{\rm E})} \, R \Big)^{-2\epsilon} ~\Big( \kappa_{(a)} + \delta_{\kappa(a)} \Big)\,,
\end{align}
where, the renormalized post-adiabatic Love number $\kappa_{(a)}$ depends on the external scale $R$, which we analyse in detail in Sec. \ref{sec_renorm_pAdterm}. 
One important thing to note is that the Lagrangian presented in Eq.~\eqref{eq_Lag_intro} can be changed by a field redefinition, which in turn redefines the different Love numbers. Hence, the above mentioned procedure of renormalizing the post-adiabatic Love number $\kappa$ is particular to this representation of the Lagrangian.

Moreover, the counterterm presented in Eq.~\eqref{eq_Lag_CT} has an intriguing form, as it turns out to be identical to the counterterms introduced to eliminate the divergences in the radiation emitted by each NS, as discussed in~\cite{Blanchet:1997jj,Goldberger:2009qd}. In Sec. \ref{sec_rad_NS}, we discuss further into the connection of this counterterm with radiation.
Similar set of counterterms was also observed in Refs.~\cite{Barack:2023oqp,Saketh:2023bul}.

\subsection{Beta-function for post-adiabatic Love number}
\label{sec_renorm_pAdterm}

Now both the gravitational constant and the post-adiabatic Love number depends on the scale $R$. However, any physical quantity should be independent of the arbitrary scale $R$. Indeed by demanding the scale independence of the bare couplings in the Lagrangian we obtain the corresponding  renormalization group equations.

For the $d$-dimensional gravitational coupling, we obtain,
\begin{align}
  0= \frac{\dd G_d}{\dd R}  = \frac{\dd}{\dd R}\left[\left(\sqrt{4\pi e^{\gamma_E}}R\right)^\epsilon G_N\right]  \,,
\end{align}
which implies a trivial beta function for $G_N$ given by,
\begin{align}
    \beta(G_N) \equiv R \, \frac{\dd G_N}{\dd R} = -\epsilon\left(\sqrt{4\pi e^{\gamma_E}}R\right)^{\epsilon}   G_N \stackrel{\epsilon\rightarrow0}{=}0 \,.
\end{align}
Similarly, for the post-adiabatic coupling, we obtain
\begin{align}
0=\frac{\dd \kappa_{d(a)}}{\dd R}   &=\frac{\dd}{\dd R}\left[ \Big(\sqrt{4 \pi \exp(\gamma_{\rm E})} \, R \Big)^{-2\epsilon} \left(\kappa_{(a)} + \delta_{\kappa_{(a)}} \right) \right] \,,
\end{align}
which results in the following beta function,
\footnote{Similar beta-function can be found for the parameter $\lambda_8$ of the following scalar theory,
\begin{align}
    \mathcal{L}= \frac12 (\partial_\mu\phi)^2 - \frac12 m^2\phi^2 - \frac{1}{6!}\lambda_6 \phi^6 - \frac{1}{8!} \lambda_6^2\lambda_8 \phi^8 \nonumber
\end{align}
}
\begin{align}
    \beta(\kappa_{(a)}) \equiv R \, \frac{\dd\kappa_{(a)}}{\dd R}&=-\frac{214}{105} \,.
\end{align}
The solution of the above given beta function is 
\begin{align}
\label{eq_RG_sol}
    \kappa_{(a)}(R)=\kappa_{(a)}(R_0)-\frac{214}{105} \log\left(\frac{R}{R_0}\right) \,,
\end{align}
where $R_0$ in the integration constant.
Now the idea for choosing $R$ will be the following \cite{Goldberger:2004jt}. In principle, we are allowed to choose the scale $R$ to whatever value we want. But as shown in Eq.~\eqref{eq_mod_kappa_R}, the $R$-dependence enters the effective Hamiltonian via $\kappa(R)$ and logarithmic terms of the form $\log\left(r/R\right)$.

When matching the point particle theory to an overarching 
theory for NSs, we should choose the scale $R$ somewhere around the cutoff of the point particle theory, i.e. $R\approx R_0\approx R_{\rm NS}$ where $R_{\rm NS}$ is some characteristic lengthscale of the NS. This will remove all the logarithmic terms from the matching procedure and we obtain a value for $\kappa(R_0\approx R_{\rm NS})$. 
Then, when we analyse the two body system, we have $R_{\rm orb}\approx r$ which is much larger than $R\approx R_{\rm NS}$, where $R_{\rm orb}$ is the characteristic lengthscale of the binary, and hence the logarithm in the Lagrangian blows up. So, in this case we will have to pick $R\approx R_{\rm orb}$, so that the logarithm is small and we get $\kappa(R_{\rm orb})$ in the effective Lagrangian. Hence the logarithmic terms plays a role only in flowing the $\kappa$ from $\kappa(R_{\rm NS})$ to $\kappa(R_{\rm orb})$ using the RG solution in Eq.~\eqref{eq_RG_sol}. 
Something peculiar to observe is that, $\kappa(R_{\rm orb})\rightarrow-\infty$ when $R_{\rm orb}\rightarrow\infty$ i.e. when the radius of the binary is large. This is acceptable because as $R_{\rm orb}\rightarrow\infty$ there are factors of $1/r^m$ with the $\kappa(R_{\rm orb})$ in the effective Lagrangian that suppress this logarithmic growth.

Using the above procedure, we will see that the observables will be independent of the arbitrary scale $R$ only after performing the matching to an overarching UV theory of NSs. An example of this can be seen in Ref.~\cite{Blanchet:1997jj}, where the authors computed the flux emitted by a binary and obtained scale dependent logarithmic terms similar to that in Eq.~\eqref{eq_RG_sol}. Such terms drop out when they include terms obtained by matching the quadrupole of the binary to the dynamics of two point particle constituents of the binary \cite{Blanchet:2013haa} and the total flux is independent of the arbitrary scale.

\subsection{Radiation from a single neutron star}
\label{sec_rad_NS}

In this section, we explore the relationship between the renormalization procedure used in the previous section and the computation of radiative observables from a single NS with a dynamical quadrupole oscillating at a frequency $\omega$. 
The counter term required to achieve a finite result in our 3PN conservative computation is the same counter term that addresses the divergences in the radiative sector as the origin of the divergent is the same in both the cases (tail-type diagrams presented in fig.~\ref{fig_div_diag}).

%
The radiative sector from the ${\cal L}_{{\rm EQ}(a)}$ term in the Lagrangian was computed in Refs.~\cite{Blanchet:1997jj,Goldberger:2009qd}.
The radiative observable was expressed in terms of a matrix element of a single graviton emitted from the NS.
The expression of these amplitudes upto two loops are given in Eq.~(38) of \cite{Goldberger:2009qd}, 
where they found that these amplitudes contain UV divergent terms and they can be removed by modifying the dynamics of the quadrupole as 
\begin{align}\label{eq_Q_renorm}
\bm{Q}_{ij}^B(\omega) = \bm{Q}_{ij}(\omega,\mu) \Bigg[1+ \frac{107}{105} \left(\frac{G_N m_{(a)} \omega}{c^3}\right)^2 \frac{1}{\epsilon}\Bigg] \,.
\end{align}
If we only consider the ${\cal L}_{{\rm EQ}(a)}$ term,
the above is equivalent to adding counter terms of the form shown in Eq. \eqref{eq_Lag_CT}. This can be seen easily as follows,
\begin{align}
- \frac{z_{(a)}}{2} \bm{E}_{(a)}^{ij}\bm{Q}_{(a)}^{{\rm B}ij}  
 &= - \frac{z_{(a)}}{2} \bm{Q}_{(a)}^{ij}   \Bigg[1+ \frac{107}{105} \left(\frac{G_N m_{(a)} }{c^3}\right)^2 \left(-\frac{\dd^2}{\dd t^2}\right) \frac{1}{\epsilon}\Bigg]  \bm{E}_{(a)}^{ij}  \nonumber\\
&
=- \frac{z_{(a)}}{2} \bm{Q}_{(a)}^{ij}  \bm{E}_{(a)}^{ij}
- \delta_{\kappa(a)} \frac{G_N^2 m_{(a)}^2}{c^6} \frac{z_{(a)}}{2} \bm{Q}_{(a)}^{ij}  \ddot{\bm{E}}_{(a)}^{ij}\,.

\end{align}
In our case, since we have a definite dynamics assigned to the quadrupole by the ${\cal L}_{\rm{FD}(a)}$ and ${\cal L}_{\rm{MQ}(a)}$ terms, the renormalization of the quadrupole as done in Refs.~\cite{Blanchet:1997jj,Goldberger:2009qd} introduces extra divergent terms. 
So, rather than performing the renormalization of the quadrupole, we renormalize the post-adiabatic Love number by adding the counter term given in Eq.~\eqref{eq_Lag_CT}, which can also remove the divergence as obtained in Eq.~(38) of \cite{Goldberger:2009qd} in the radiative sector .

Additionally, in Ref.~\cite{Goldberger:2004jt}, it was observed that when starting with the point particle action (non-spinning and without tides), the stress-energy tensor and, consequently, the metric produced by a single compact object contain divergent terms. When we extend this analysis to a binary system comprising two such point particles, the divergent metric generated by the first object affects the second object, leading to the emergence of divergences in the two-body interaction potential.
The divergent terms from both the metric and the interaction potential were removed with the introduction of a counter term in the point particle action and ultimately these counter terms vanished by applying a field redefinition in the Lagrangian. We see an analogous case for the 3PN tidal interaction potential. The radiative sector 
from the ${\cal L}_{{\rm EQ}(a)}$ term in the Lagrangian contains divergent pieces. If one now computes the metric from this, it should also be divergent. Hence we should expect the same divergence to also show up in the two body tidal interaction potential. The only difference being, the counter term that removes the tidal divergence, cannot be removed by a field redefinition as opposed to the same in the point particle sector.
\section{Dynamical tides}
\label{sec_results}
In this section, we present the result of the effective two-body Hamiltonian with dynamical gravitoelectric tides. We compute the Hamiltonian $\mathcal{H}$ from the Lagrangian obtained in the previous section using a Legendre transformation
\begin{align}\label{eq_ham_def}
\mathcal{H}(\bm{x},\bm{p},\bm{Q})= \sum_{a=1,2}\left( \, \bm{p}_{(a)}^i {\bm v}_{(a)}^i  +\bm{P}_{(a)}^{ij}\dot{\bm{Q}}_{(a)}^{ij} \, \right) - \mathcal{L}(\bm{x},{\bm v},\bm{Q}) \ .
\end{align}
We express this Hamiltonian in terms of the total mass of the binary, denoted by $M=m_{(1)}+m_{(2)}$, the reduced mass by $\mu=m_{(1)}m_{(2)}/M$, the mass ratio by $q=m_{(1)}/m_{(2)}$, the symmetric mass ratio $\nu=\mu/M$, and the antisymmetric mass ratio $\delta=(m_{(1)}-m_{(2)})/M$, which are related to each other by,
\begin{align}
\nu=\frac{m_{(1)} m_{(2)}}{M^2}=\frac{\mu}{M}=\frac{q}{(1+q)^2} = \frac{(1-\delta^2)}{4} \,.
\end{align}
We express the results in the center-of-mass (COM) frame of reference and define the momentum in the COM frame as $\bm{p} \equiv \bm{p}_{(1)}=-\bm{p}_{(2)}$. In the COM frame, the orbital angular momentum is defined as $\bm{L} = \bm{r}\times \bm{p}$. Hence, we can write $p^2 = p_r^2 + L^2 / r^2$, where $p_r=\bm{p} \cdot \bm{n}$, $p = |\bm{p}|$ and $L = |\bm{L}|$.
We write the Hamiltonian in terms of dimensionless quantities, which are obtained by rescaling all the variables as follows
\begin{align}
&\widetilde{\bm{p}}=\frac{1}{c}\frac{\bm{p}}{\mu} \, ,\quad
\widetilde{\bm{r}}=\frac{c^2}{G_N} \frac{\bm{r}}{M} \, ,\quad
\widetilde{\bm{L}}=\frac{c}{G_N} \frac{\bm{L}}{M \mu} \, ,\quad 
\widetilde{\mathcal{H}}=\frac{1}{c^2} \frac{\mathcal{H}}{\mu} \, , \quad \widetilde{\lambda}= \frac{c^{10}}{G_N^4}\frac{\lambda}{M^5} \,,\nonumber \\
\label{eq_rescale}
&\widetilde{\bm{Q}}_{(a)}=\frac{c^4}{G_N^2} \frac{\bm{Q}_{(a)}}{M^2 \mu} \, ,\quad
\widetilde{\bm{S}}_{Q(a)}=\frac{c}{G_N} \frac{\bm{S}_{Q(a)}}{M \mu} \,, \quad \text{and} \quad
\widetilde{\bm{M}}_{Q(a)}=\frac{1}{c^2}\frac{\bm{M}_{Q(a)}}{\mu} \,.
\end{align}
The total EFT Hamiltonian in the dimensionless parameters is given by
\begin{align}\label{eq_ham_dy}
\widetilde{\mathcal{H}} = \widetilde{\mathcal{H}}_{\rm pp} + \widetilde{\mathcal{H}}_{\rm DT}\ ,
\end{align}
where
\begin{subequations}
\begin{align}\label{eq_Ham_pp}
\widetilde{\mathcal{H}}_{\rm pp} &= \widetilde{\mathcal{H}}_{\text{0PN}}+\left(\frac{1}{c^2}\right)\widetilde{\mathcal{H}}_{\text{1PN}}+\left(\frac{1}{c^4}\right)\widetilde{\mathcal{H}}_{\text{2PN}}
+\left(\frac{1}{c^6}\right)\widetilde{\mathcal{H}}_{\text{3PN}}
+\mathcal{O}\left(\frac{1}{c^8}\right) \,, \\
\widetilde{{\cal H}}_{\rm DT} &= \sum_{n = 0}^{3}
\left(\frac{1}{c^2}\right)^n \, 
\left(
\widetilde{{\cal H}}^{\rm EQ}_{n{\rm PN}}
+ \widetilde{{\cal H}}^{\rm FD}_{n{\rm PN}}
+ \widetilde{{\cal H}}^{\rm MQ}_{n{\rm PN}}
\right)
+ \left(\frac{1}{c^6}\right) \widetilde{\mathcal{H}}^{\rm \ddot{E}Q}_{\rm 3PN}
+\mathcal{O}\left(\frac{1}{c^8}\right) \,.
\end{align}
\end{subequations}
The point particle Hamiltonian till 3PN is presented in the same gauge in Appendix C.1 of Ref.~\cite{Mandal:2022nty}, the tidal Hamiltonian up to 2PN is presented in Ref.~\cite{Mandal:2023lgy} and the novel result of 3PN Hamiltonian is given as,
\begin{subequations}
\label{eq:new_3PN_H}
\begin{align} 
\widetilde{\mathcal{H}}^{\rm EQ + \ddot{E}Q}_{\rm 3PN}
=& 
\left( \widetilde{\bm{Q}}^{ij}_{(1)}\widetilde{\bm{r}}^{\,i}\widetilde{\bm{r}}^{\,j} \right) \Bigg\{
\widetilde{L}^6 \left(-\frac{15 \nu ^4}{32 \widetilde{r}^{11}}-\frac{135 \nu ^3}{32 \widetilde{r}^{11}}+\frac{45 \nu ^2}{32 \widetilde{r}^{11}}\right)
+\widetilde{L}^4 \left(\widetilde{p}_r^2 \left(-\frac{15 \nu ^4}{8 \widetilde{r}^9}-\frac{585 \nu ^3}{32 \widetilde{r}^9}+\frac{327 \nu ^2}{32 \widetilde{r}^9}\right)\right. \nonumber\\
&\left.  +\frac{75 \nu ^3}{4 \widetilde{r}^{10}}+\frac{137 \nu ^2}{4 \widetilde{r}^{10}}-\frac{101 \nu }{32 \widetilde{r}^{10}}\right)+\widetilde{L}^2 \left[\widetilde{p}_r^4 \left(-\frac{51 \nu ^4}{16 \widetilde{r}^7}-\frac{225 \nu ^3}{8 \widetilde{r}^7}+\frac{1527 \nu ^2}{32 \widetilde{r}^7}-\frac{105 \nu }{8 \widetilde{r}^7}\right) 
\right.  \nonumber\\
&\left. 
+ \widetilde{p}_r^2 \left(\frac{1977 \nu ^3}{32 \widetilde{r}^8}-\frac{383 \nu ^2}{16 \widetilde{r}^8}-\frac{79 \nu }{32 \widetilde{r}^8}\right)+\frac{\left(-3763200 ~\bar{\kappa}_{(1)}+77175 \pi ^2-35396224\right) \nu ^2}{501760 \widetilde{r}^9} \right. \nonumber\\
&\left. +\frac{(3675 ~\bar{\kappa}_{(1)}+16651) \nu }{490 \widetilde{r}^9}-\frac{1269 \nu ^3}{112 \widetilde{r}^9}\right] +\widetilde{p}_r^2 \left(\frac{3 \left(3763200 ~\bar{\kappa}_{(1)}-77175 \pi ^2+11490272\right) \nu ^2}{627200 \widetilde{r}^7} \right. \nonumber\\
&\left.-\frac{3 (117600 ~\bar{\kappa}_{(1)}+679671) \nu }{19600 \widetilde{r}^7}-\frac{153 \nu ^3}{7 \widetilde{r}^7}\right)+\widetilde{p}_r^4 \left(\frac{477 \nu ^3}{32 \widetilde{r}^6}-\frac{291 \nu ^2}{8 \widetilde{r}^6}+\frac{327 \nu }{32 \widetilde{r}^6}\right)  \nonumber\\
& +\widetilde{p}_r^6\left(-\frac{6 \nu ^4}{\widetilde{r}^5}+\frac{45 \nu ^3}{2 \widetilde{r}^5}-\frac{135 \nu ^2}{8 \widetilde{r}^5}+\frac{105 \nu }{32 \widetilde{r}^5}\right)+\frac{\left(705600 ~\bar{\kappa}_{(1)}-1157625 \pi ^2+14115256\right) \nu ^2}{156800 \widetilde{r}^8} \nonumber\\
&-\frac{3 (14700 ~\bar{\kappa}_{(1)}+44197) \nu }{9800 \widetilde{r}^8}  +\frac{1}{q} \Bigg[\widetilde{L}^6 \left(-\frac{15 \nu ^4}{32 \widetilde{r}^{11}}-\frac{45 \nu ^3}{16 \widetilde{r}^{11}}+\frac{105 \nu ^2}{32 \widetilde{r}^{11}}-\frac{21 \nu }{32 \widetilde{r}^{11}}\right)\nonumber\\
&+\widetilde{L}^4 \left(\widetilde{p}_r^2 \left(-\frac{15 \nu ^4}{8 \widetilde{r}^9}-\frac{399 \nu ^3}{32 \widetilde{r}^9}+\frac{159 \nu ^2}{32 \widetilde{r}^9}-\frac{39 \nu }{32 \widetilde{r}^9}\right)+\frac{591 \nu ^3}{32 \widetilde{r}^{10}}+\frac{973 \nu ^2}{32 \widetilde{r}^{10}}-\frac{15 \nu }{2 \widetilde{r}^{10}}\right)\nonumber\\
&+\widetilde{L}^2 \left(\widetilde{p}_r^4 \left(-\frac{51 \nu ^4}{16 \widetilde{r}^7}-\frac{321 \nu ^3}{16 \widetilde{r}^7}+\frac{399 \nu ^2}{32 \widetilde{r}^7}-\frac{3 \nu }{32 \widetilde{r}^7}\right) +\widetilde{p}_r^2 \left(\frac{1977 \nu ^3}{32 \widetilde{r}^8}+\frac{933 \nu ^2}{32 \widetilde{r}^8}-\frac{6 \nu }{\widetilde{r}^8}\right) \right.\nonumber\\
& \left. +\frac{\left(77175 \pi ^2-1536 (2450 ~\bar{\kappa}_{(1)}+28169)\right) \nu ^2}{501760 \widetilde{r}^9}-\frac{1269 \nu ^3}{112 \widetilde{r}^9}-\frac{369 \nu }{8 \widetilde{r}^9}\right)\nonumber\\
& \left. +\widetilde{p}_r^2 \left(-\frac{3 \left(-3763200 ~\bar{\kappa}_{(1)}+77175 \pi ^2-659872\right) \nu ^2}{627200 \widetilde{r}^7}-\frac{153 \nu ^3}{7 \widetilde{r}^7}+\frac{117 \nu }{8 \widetilde{r}^7}\right)\right.\nonumber\\
&\left.+\widetilde{p}_r^4 \left(\frac{219 \nu ^3}{8 \widetilde{r}^6}-\frac{735 \nu ^2}{32 \widetilde{r}^6}+\frac{9 \nu }{2 \widetilde{r}^6}\right)+\widetilde{p}_r^6 \left(-\frac{6 \nu ^4}{\widetilde{r}^5}+\frac{27 \nu ^3}{2 \widetilde{r}^5}-\frac{45 \nu ^2}{8 \widetilde{r}^5}+\frac{15 \nu }{32 \widetilde{r}^5}\right)\right.\nonumber\\
&\left.+\frac{\left(705600 ~\bar{\kappa}_{(1)}-1157625 \pi ^2+15841456\right) \nu ^2}{156800 \widetilde{r}^8}+\frac{63 \nu }{2 \widetilde{r}^8}\Bigg]
\right\}\nonumber\\
&+ \left(\widetilde{\bm{Q}}^{ij}_{(1)}\widetilde{\bm{L}}^i\widetilde{\bm{L}}^j\right) \left\{\widetilde{p}_r^4 \left(-\frac{21 \nu ^4}{16 \widetilde{r}^5}-\frac{45 \nu ^3}{4 \widetilde{r}^5}+\frac{24 \nu ^2}{\widetilde{r}^5}-\frac{105 \nu }{16 \widetilde{r}^5}\right)+
\widetilde{p}_r^2 \left(\frac{225 \nu ^3}{16 \widetilde{r}^6}+\frac{577 \nu ^2}{8 \widetilde{r}^6}+\frac{77 \nu }{4 \widetilde{r}^6}\right)\right.\nonumber\\
& \left.+\widetilde{L}^4\left(\frac{15 \nu ^2}{16 \widetilde{r}^9}-\frac{45 \nu ^3}{16 \widetilde{r}^9}\right)+\widetilde{L}^2\left(\widetilde{p}_r^2 \left(-\frac{3 \nu ^4}{8 \widetilde{r}^7}-\frac{45 \nu ^3}{4 \widetilde{r}^7}+\frac{21 \nu ^2}{4 \widetilde{r}^7}\right)+\frac{9 \nu ^3}{2 \widetilde{r}^8}+\frac{403 \nu ^2}{16 \widetilde{r}^8}-\frac{209 \nu }{16 \widetilde{r}^8}\right)\right.\nonumber\\
& \left.+\frac{\left(3763200 ~\bar{\kappa}_{(1)}-77175 \pi ^2-25954688\right) \nu ^2}{1254400 \widetilde{r}^7}-\frac{(3675 ~\bar{\kappa}_{(1)}+52288) \nu }{1225 \widetilde{r}^7}-\frac{3 \nu ^3}{2 \widetilde{r}^7} \right. \nonumber\\
&\left. +\frac{1}{q} \left[\widetilde{L}^4 \left(-\frac{45 \nu ^3}{16 \widetilde{r}^9}+\frac{45 \nu ^2}{16 \widetilde{r}^9}-\frac{9 \nu }{16 \widetilde{r}^9}\right)+\widetilde{L}^2 \left(\widetilde{p}_r^2 \left(-\frac{3 \nu ^4}{8 \widetilde{r}^7}-\frac{12 \nu ^3}{\widetilde{r}^7}+\frac{39 \nu ^2}{8 \widetilde{r}^7}-\frac{3 \nu }{4 \widetilde{r}^7}\right)\right.\right.\right.\nonumber\\
&\left.+\frac{9 \nu ^3}{2 \widetilde{r}^8}+\frac{325 \nu ^2}{16 \widetilde{r}^8}-\frac{6 \nu }{\widetilde{r}^8}\right)+\widetilde{p}_r^4 \left(-\frac{21 \nu ^4}{16 \widetilde{r}^5}-\frac{237 \nu ^3}{16 \widetilde{r}^5}+\frac{69 \nu ^2}{8 \widetilde{r}^5}-\frac{3 \nu }{16 \widetilde{r}^5}\right)\nonumber\\
&\left.\left.+\widetilde{p}_r^2 \left(\frac{15 \nu ^3}{\widetilde{r}^6}+\frac{1065 \nu ^2}{16 \widetilde{r}^6}-\frac{3 \nu }{\widetilde{r}^6}\right)+\frac{\left(384 (9800 ~\bar{\kappa}_{(1)}-56157)-77175 \pi ^2\right) \nu ^2}{1254400 \widetilde{r}^7}-\frac{3 \nu ^3}{2 \widetilde{r}^7}-\frac{123 \nu }{4 \widetilde{r}^7}\right]
\right\} \nonumber\\
& + \left( \widetilde{\bm{Q}}^{ij}_{(1)}\widetilde{\bm{r}}^{\,i} \widetilde{\bm{L}}^j \right)\widetilde{p}_r 
\Bigg\{\widetilde{L}^4 \left(\frac{15 \nu ^4}{16 \widetilde{r}^9}+\frac{45 \nu ^3}{16 \widetilde{r}^9}-\frac{9 \nu ^2}{16 \widetilde{r}^9}\right)+
\widetilde{L}^2 \left(\widetilde{p}_r^2 \left(\frac{63 \nu ^4}{16 \widetilde{r}^7}+\frac{45 \nu ^3}{8 \widetilde{r}^7}-\frac{45 \nu ^2}{8 \widetilde{r}^7}\right)\right.\nonumber\\
&\left.-\frac{105 \nu ^3}{8 \widetilde{r}^8}+\frac{5 \nu ^2}{2 \widetilde{r}^8}-\frac{\nu }{2 \widetilde{r}^8}\right)+\widetilde{p}_r^4 \left(\frac{123 \nu ^4}{16 \widetilde{r}^5}-\frac{45 \nu ^3}{2 \widetilde{r}^5}+\frac{129 \nu ^2}{16 \widetilde{r}^5}\right)+\widetilde{p}_r^2 \left(-\frac{81 \nu ^3}{8 \widetilde{r}^6}-\frac{229 \nu ^2}{2 \widetilde{r}^6}+\frac{453 \nu }{8 \widetilde{r}^6}\right)\nonumber\\
&\left.+\frac{\left(77175 \pi ^2-384 (9800 ~\bar{\kappa}_{(1)}+8103)\right) \nu ^2}{156800 \widetilde{r}^7}+\frac{(235200 ~\bar{\kappa}_{(1)}+2662147) \nu }{9800 \widetilde{r}^7}+\frac{93 \nu ^3}{7 \widetilde{r}^7} \right. \nonumber\\
& +\frac{1}{q} \left[  \widetilde{L}^4 \left(\frac{15 \nu ^4}{16 \widetilde{r}^9}+\frac{3 \nu ^3}{\widetilde{r}^9}-\frac{33 \nu ^2}{16 \widetilde{r}^9}+\frac{3 \nu }{16 \widetilde{r}^9}\right)+
\widetilde{L}^2 \left(\widetilde{p}_r^2 \left(\frac{63 \nu ^4}{16 \widetilde{r}^7}+\frac{147 \nu ^3}{16 \widetilde{r}^7}-\frac{45 \nu ^2}{8 \widetilde{r}^7}+\frac{9 \nu }{8 \widetilde{r}^7}\right)\right.\right.\nonumber\\
&\left.\left.-\frac{15 \nu ^3}{\widetilde{r}^8}+\frac{3 \nu ^2}{8 \widetilde{r}^8}+\frac{3 \nu }{\widetilde{r}^8}\right)+\widetilde{p}_r^4 \left(\frac{123 \nu ^4}{16 \widetilde{r}^5}-\frac{81 \nu ^3}{16 \widetilde{r}^5}-\frac{57 \nu ^2}{16 \widetilde{r}^5}+\frac{15 \nu }{16 \widetilde{r}^5}\right)+\widetilde{p}_r^2 \left(-\frac{171 \nu ^3}{8 \widetilde{r}^6}-\frac{555 \nu ^2}{8 \widetilde{r}^6}+\frac{9 \nu }{\widetilde{r}^6}\right)\right.\nonumber\\
&\left.\left.+\frac{\left(-3763200 ~\bar{\kappa}_{(1)}+77175 \pi ^2-5256352\right) \nu ^2}{156800 \widetilde{r}^7}+\frac{93 \nu ^3}{7 \widetilde{r}^7}+\frac{30 \nu }{\widetilde{r}^7} \right] \right\}
\nonumber\\
&+(1\leftrightarrow 2) \,,
\end{align} 
\begin{align} 
\widetilde{\mathcal{H}}^{\rm FD}_{\rm 3PN} =&
\left( \widetilde{\bm{S}}_{Q(1)}\cdot \widetilde{\bm{L}} \right)  \Bigg\{
\widetilde{L}^4 \left(\frac{11 \nu ^3}{8 \widetilde{r}^7}+\frac{9 \nu ^2}{16 \widetilde{r}^7}-\frac{\nu }{4 \widetilde{r}^7}\right)+\widetilde{L}^2 \left(\widetilde{p}_r^2 \left(\frac{23 \nu ^3}{4 \widetilde{r}^5}+\frac{21 \nu ^2}{8 \widetilde{r}^5}-\frac{2 \nu }{\widetilde{r}^5}\right)-\frac{17 \nu ^3}{16 \widetilde{r}^6} \right.\nonumber\\
&\left.-\frac{275 \nu ^2}{16 \widetilde{r}^6}-\frac{17 \nu }{4 \widetilde{r}^6}\right)+\widetilde{p}_r^2 \left(-\frac{63 \nu ^3}{16 \widetilde{r}^4}-\frac{207 \nu ^2}{8 \widetilde{r}^4}+\frac{9 \nu }{2 \widetilde{r}^4}\right)+\widetilde{p}_r^4 \left(\frac{10 \nu ^3}{\widetilde{r}^3}-\frac{81 \nu ^2}{8 \widetilde{r}^3}+\frac{2 \nu }{\widetilde{r}^3}\right) +\frac{75 \nu ^2}{8 \widetilde{r}^5}+\frac{25 \nu }{2 \widetilde{r}^5}  \nonumber\\
& \left. +\frac{1}{q} \left[ \widetilde{L}^4 \left(\frac{17 \nu ^3}{16 \widetilde{r}^7}-\frac{9 \nu ^2}{4 \widetilde{r}^7}+\frac{7 \nu }{16 \widetilde{r}^7}\right)+\widetilde{L}^2 \left(\widetilde{p}_r^2 \left(\frac{73 \nu ^3}{16 \widetilde{r}^5}-\frac{3 \nu ^2}{\widetilde{r}^5}+\frac{7 \nu }{8 \widetilde{r}^5}\right)-\frac{17 \nu ^3}{16 \widetilde{r}^6}-\frac{133 \nu ^2}{16 \widetilde{r}^6}+\frac{27 \nu }{8 \widetilde{r}^6}\right) \right.\right.\nonumber\\
&\left. +\widetilde{p}_r^2 \left(-\frac{63 \nu ^3}{16 \widetilde{r}^4}-\frac{411 \nu ^2}{16 \widetilde{r}^4}+\frac{27 \nu }{8 \widetilde{r}^4}\right)+\widetilde{p}_r^4 \left(\frac{131 \nu ^3}{16 \widetilde{r}^3}-\frac{9 \nu ^2}{2 \widetilde{r}^3}+\frac{7 \nu }{16 \widetilde{r}^3}\right) +\frac{43 \nu ^2}{8 \widetilde{r}^5}+\frac{21 \nu }{2 \widetilde{r}^5} \right]
\Bigg\} \nonumber\\
& +(1\leftrightarrow 2)  \,,
\end{align} 
\begin{align} 
\widetilde{\mathcal{H}}^{\rm MQ}_{\text{3PN}} =& \widetilde{\bm{M}}_{Q(1)}
\left\{\widetilde{L}^6 \left(\frac{5 \nu ^2}{16 \widetilde{r}^6}-\frac{5 \nu ^3}{8 \widetilde{r}^6}\right)+\widetilde{L}^4 \left(\widetilde{p}_r^2 \left(\frac{15 \nu ^2}{16 \widetilde{r}^4}-\frac{15 \nu ^3}{8 \widetilde{r}^4}\right)+\frac{3 \nu ^3}{8 \widetilde{r}^5}+\frac{\nu ^2}{\widetilde{r}^5}\right)\right.\nonumber\\
&\left.+\widetilde{L}^2 \left(\widetilde{p}_r^2 \left(\frac{\nu ^3}{\widetilde{r}^3}+\frac{\nu ^2}{4 \widetilde{r}^3}+\frac{\nu }{\widetilde{r}^3}\right)+\widetilde{p}_r^4 \left(\frac{15 \nu ^2}{16 \widetilde{r}^2}-\frac{15 \nu ^3}{8 \widetilde{r}^2}\right)+\frac{3 \nu ^2}{4 \widetilde{r}^4}+\frac{49 \nu }{8 \widetilde{r}^4}\right)+\widetilde{p}_r^2 \left(\frac{7 \nu ^2}{4 \widetilde{r}^2}+\frac{10 \nu }{\widetilde{r}^2}\right) \right.\nonumber\\
&\left.+\widetilde{p}_r^4 \left(\frac{\nu ^3}{\widetilde{r}}+\frac{5 \nu }{8 \widetilde{r}}\right)+\left(\frac{5 \nu ^2}{16}-\frac{5 \nu ^3}{8}\right) \widetilde{p}_r^6 +\frac{3 \nu ^2}{4 \widetilde{r}^3}+\frac{\nu }{2 \widetilde{r}^3} +\frac{1}{q} \left[\widetilde{L}^6 \left(-\frac{15 \nu ^3}{16 \widetilde{r}^6}+\frac{5 \nu ^2}{4 \widetilde{r}^6}-\frac{5 \nu }{16 \widetilde{r}^6}\right) \right.\right. \nonumber\\
& \left.\left. +\widetilde{L}^4 \left(\widetilde{p}_r^2 \left(-\frac{45 \nu ^3}{16 \widetilde{r}^4}+\frac{15 \nu ^2}{4 \widetilde{r}^4}-\frac{15 \nu }{16 \widetilde{r}^4}\right)+\frac{3 \nu ^3}{8 \widetilde{r}^5}+\frac{45 \nu ^2}{8 \widetilde{r}^5}-\frac{15 \nu }{8 \widetilde{r}^5}\right) +\widetilde{L}^2 \left(\widetilde{p}_r^2 \left(\frac{\nu ^3}{\widetilde{r}^3}+\frac{41 \nu ^2}{4 \widetilde{r}^3}-\frac{15 \nu }{4 \widetilde{r}^3}\right) \right.\right. \right.\nonumber\\
&\left.\left.\left.+\widetilde{p}_r^4 \left(-\frac{45 \nu ^3}{16 \widetilde{r}^2}+\frac{15 \nu ^2}{4 \widetilde{r}^2}-\frac{15 \nu }{16 \widetilde{r}^2}\right)-\frac{23 \nu ^2}{8 \widetilde{r}^4}-\frac{11 \nu }{4 \widetilde{r}^4}\right) +\widetilde{p}_r^2 \left(-\frac{13 \nu ^2}{4 \widetilde{r}^2}-\frac{9 \nu }{4 \widetilde{r}^2}\right) \right.\right.\nonumber\\
&\left.\left. +\widetilde{p}_r^6\left(-\frac{15 \nu ^3}{16}+\frac{5 \nu ^2}{4}-\frac{5 \nu }{16}\right) +\widetilde{p}_r^4 \left(\frac{\nu ^3}{\widetilde{r}}+\frac{5 \nu ^2}{\widetilde{r}}-\frac{15 \nu }{8 \widetilde{r}}\right) +\frac{3 \nu ^2}{4 \widetilde{r}^3}-\frac{\nu }{2 \widetilde{r}^3} \right]
\right\} \nonumber\\
& +(1\leftrightarrow 2) \,.
\end{align}
\end{subequations}
where,
\begin{align} \label{eq_mod_kappa_R}
    \bar{\kappa}_{(1)}(R) = \kappa_{(1)}(R) - \frac{214}{105} \log \left(\frac{r}{R}\right)\,,
\end{align}
that combines the terms that depend on the external length scale $R$. The effective Hamiltonian in a general reference frame is provided in the ancillary file \texttt{Hamiltonian-DT.m}.

\section{Adiabatic tides}
\label{sec_adiabatic}

In this section, we present our results for the adiabatic tides, that is, we take the $\omega_f\rightarrow\infty$ limit of the Hamiltonian~\eqref{eq_ham_dy}. This eliminates the dependence of Hamiltonian on the variables $\bm{Q}_{(a)}$, $\bm{S}_{Q(a)}$ and $\bm{M}_{Q(a)}$, and hence simplifies our further calculations. We then compute the binding energy and scattering angle using the adiabatic Hamiltonian and compare these against known results in the literature.

The adiabatic limit physically refers to the quadrupole mode being locked to the external tidal field induced by the binary companion.
In this case, the equation of motion for the $\bm{Q}_{(a)}^{ij}$ is given by
\begin{align}
\label{eq_QtoE}
    \bm{Q}^{ij}_{(a)}=-\lambda_{(a)} \bm{E}^{ij}_{(a)}- \lambda_{(a)}\kappa_{d(a)} \frac{G_d^2 m_{(a)}^2}{c^6} \ddot{\bm{E}}^{ij}_{(a)}\,.
\end{align}
We can then substitute Eqs.~\eqref{eq_QtoE} in the Hamiltonian~\eqref{eq_ham_dy} to obtain the effective Hamiltonian for adiabatic tides. 

Computationally it is efficient to compute dynamic tides and then take the adiabatic limit, than directly computing the adiabatic tides from the Lagrangian~\eqref{eq_Lag_ad}. This is because all the Feynman diagrams generated by \eqref{eq_Lag_ad} will be factorizable due to $E^2$ terms. Hence in this case we can compute 4-loop ($G_N^5$) terms in the adiabatic observables by doing a 3-loop computation with the dynamic tides.
The counter term for an adiabatic calculation starting from Lagrangian given in Eq.~\eqref{eq_Lag_ad}, can be obtained by using Eq.~\eqref{eq_kappa_delta} in Eq.~\eqref{eq_Lag_ad}.

\subsection{Effective Hamiltonian}
The effective adiabatic Hamiltonian is given by\footnote{At leading order, adiabatic tides contributes at 5PN which can be easily seen writing the Hamiltonian in the form of the dimensional Love number as shown in Eq.~\eqref{eq_rescale}. We show here the relative scaling with respect to the 5PN order contribution.},
\begin{align}\label{eq_ham_ad}
\widetilde{\mathcal{H}} = \widetilde{\mathcal{H}}_{\rm pp} + \widetilde{\mathcal{H}}_{\rm AT}\ ,
\end{align}
where
\begin{subequations}
\begin{align}
\widetilde{\mathcal{H}}_{\rm pp} &= \widetilde{\mathcal{H}}_{\text{0PN}}+\left(\frac{1}{c^2}\right)\widetilde{\mathcal{H}}_{\text{1PN}}+\left(\frac{1}{c^4}\right)\widetilde{\mathcal{H}}_{\text{2PN}}
++\left(\frac{1}{c^6}\right)\widetilde{\mathcal{H}}_{\text{3PN}}
+\mathcal{O}\left(\frac{1}{c^8}\right) \ , \\
\widetilde{\mathcal{H}}_{\rm AT} 
&= \widetilde{\mathcal{H}}^{\text{AT}}_{\text{0PN}}+\left(\frac{1}{c^{2}}\right)\widetilde{\mathcal{H}}^{\text{AT}}_{\text{1PN}}+\left(\frac{1}{c^{4}}\right)\widetilde{\mathcal{H}}^{\text{AT}}_{\text{2PN}}
+\left(\frac{1}{c^6}\right)\widetilde{\mathcal{H}}^{\text{AT}}_{\text{3PN}}
+\mathcal{O}\left(\frac{1}{c^{8}}\right) \ .
\end{align}
\end{subequations}
The adiabatic Hamiltonian up to 2PN is presented in \cite{Mandal:2023lgy} and the novel result of 3PN Hamiltonian is given as,
\begin{align}
    \widetilde{{\cal H}}^{\rm AT}_{3{\rm PN}} =& \widetilde{\lambda}_{(1)} \left\{
\widetilde{L}^6 \left(-\frac{45 \nu ^3}{32 \widetilde{r}^{12}}-\frac{15 \nu ^2}{4 \widetilde{r}^{12}}-\frac{33 \nu }{32 \widetilde{r}^{12}}\right)
+\widetilde{L}^4 \left(\widetilde{p}_r^2 \left(-\frac{189 \nu ^3}{32 \widetilde{r}^{10}}+\frac{99 \nu ^2}{8 \widetilde{r}^{10}}+\frac{711 \nu }{32 \widetilde{r}^{10}}\right)-\frac{9 \nu ^2}{\widetilde{r}^{11}}+\frac{15 \nu }{4 \widetilde{r}^{11}}\right)\right. \nonumber\\
&+\widetilde{L}^2 \left[\widetilde{p}_r^2 \left(-\frac{9 \nu ^3}{2 \widetilde{r}^9}-\frac{2775 \nu ^2}{16 \widetilde{r}^9}-\frac{72 \nu }{\widetilde{r}^9}\right)+\widetilde{p}_r^4 \left(-\frac{99 \nu ^3}{32 \widetilde{r}^8}+\frac{108 \nu ^2}{\widetilde{r}^8}-\frac{1431 \nu }{32 \widetilde{r}^8}\right) \right.
 \nonumber\\
& \left. +\frac{3 (117600 ~\bar{\kappa}_{(1)}+1584771) \nu }{19600 \widetilde{r}^{10}}+\frac{93 \nu ^2}{2 \widetilde{r}^{10}}\right] +\widetilde{p}_r^2 \left(\frac{9993 \nu ^2}{56 \widetilde{r}^8}-\frac{3 (117600 ~\bar{\kappa}_{(1)}+626121) \nu }{9800 \widetilde{r}^8}\right) \nonumber\\
&+\widetilde{p}_r^4 \left(-\frac{45 \nu ^3}{\widetilde{r}^7}-\frac{729 \nu ^2}{8 \widetilde{r}^7}+\frac{777 \nu }{16 \widetilde{r}^7}\right) +\widetilde{p}_r^6 \left(\frac{1485 \nu ^3}{32 \widetilde{r}^6}-\frac{465 \nu ^2}{8 \widetilde{r}^6}+\frac{465 \nu }{32 \widetilde{r}^6}\right) -\frac{3 (29400 ~\bar{\kappa}_{(1)}+429119) \nu }{9800 \widetilde{r}^9} \nonumber\\
& +\frac{1}{q} \Bigg[\widetilde{L}^6 \left(-\frac{15 \nu ^3}{8 \widetilde{r}^{12}}-\frac{75 \nu ^2}{8 \widetilde{r}^{12}}-\frac{99 \nu }{16 \widetilde{r}^{12}}+\frac{33}{32 \widetilde{r}^{12}}\right)+\widetilde{L}^4 \left(\widetilde{p}_r^2 \left(-\frac{9 \nu ^3}{\widetilde{r}^{10}}-\frac{171 \nu ^2}{8 \widetilde{r}^{10}}-\frac{27 \nu }{16 \widetilde{r}^{10}}+\frac{9}{32 \widetilde{r}^{10}}\right)\right.\nonumber\\
&\left.+\frac{273 \nu ^2}{16 \widetilde{r}^{11}}+\frac{1545 \nu }{16 \widetilde{r}^{11}}+\frac{495}{16 \widetilde{r}^{11}}\right)+\widetilde{L}^2 \left(\widetilde{p}_r^4 \left(-\frac{99 \nu ^3}{8 \widetilde{r}^8}+\frac{315 \nu ^2}{8 \widetilde{r}^8}+\frac{27 \nu }{16 \widetilde{r}^8}-\frac{9}{32 \widetilde{r}^8}\right) \right.\nonumber\\
&\left. +\widetilde{p}_r^2 \left(-\frac{9 \nu ^3}{2 \widetilde{r}^9}-\frac{831 \nu ^2}{16 \widetilde{r}^9}+\frac{567 \nu }{8 \widetilde{r}^9}-\frac{99}{8 \widetilde{r}^9}\right)+\frac{867 \nu ^2}{28 \widetilde{r}^{10}}+\frac{3 \left(8384+63 \pi ^2\right) \nu }{512 \widetilde{r}^{10}}-\frac{1335}{8 \widetilde{r}^{10}}\right)\nonumber\\
& \left. +\widetilde{p}_r^2 \left(\frac{2097 \nu ^2}{16 \widetilde{r}^8}-\frac{3 \left(20576+63 \pi ^2\right) \nu }{256 \widetilde{r}^8}+\frac{261}{8 \widetilde{r}^8}\right)+\widetilde{p}_r^4 \left(-\frac{45 \nu ^3}{\widetilde{r}^7}-\frac{105 \nu ^2}{8 \widetilde{r}^7}-\frac{327 \nu }{16 \widetilde{r}^7}+\frac{99}{16 \widetilde{r}^7}\right)\right.\nonumber\\
&\left.+\widetilde{p}_r^6 \left(\frac{99 \nu ^3}{4 \widetilde{r}^6}-\frac{69 \nu ^2}{8 \widetilde{r}^6}-\frac{45 \nu }{16 \widetilde{r}^6}+\frac{15}{32 \widetilde{r}^6}\right) -\frac{1599 \nu ^2}{56 \widetilde{r}^9}+\frac{\left(6376-945 \pi ^2\right) \nu }{64 \widetilde{r}^9}+\frac{519}{4 \widetilde{r}^9}\Bigg]
\right\}\nonumber\\
& +(1\leftrightarrow 2) \,.
\end{align}
This Hamiltonian in a general reference frame is provided by us in the ancillary file \texttt{Hamiltonian-AT.m}.
Following the procedure described in Sec.~5.2 of Ref.~\cite{Mandal:2023lgy}, we have validated the above result by computing the complete Poincare algebra \cite{Damour:2000kk,Levi:2016ofk} using the above result. The expression of the center of mass $\bm{G}^i$ is given in the ancilary file \texttt{Poincare\_Algebra.m}.

\subsection{Binding energy}
\label{sec_binding_energy}
In this section, we compute the binding energy in the COM frame for circular orbits. The gauge invariant relation between the binding energy and the orbital frequency for circular orbits is obtained by eliminating the dependence on the radial coordinate.
For circular orbits we have
    $\partial \widetilde{\mathcal{H}}(\widetilde{r},\widetilde{L})/ \partial \widetilde{r}=0\, .
    $
%
We invert this relation to express $\widetilde{r}$ as a function of $\widetilde{L}$. 
Then, we substitute $\widetilde{L}$, written as a function of the orbital frequency $\widetilde{\omega} = \partial \widetilde{\mathcal{H}}(\widetilde{L})/\partial \widetilde{L}$, in the Hamiltonian~\eqref{eq_ham_ad}.
Following this procedure we obtain the binding energy $E$ as,
\begin{align}
    E_{\text{AT}} = E^{\rm 0PN}_{\text{AT}} + E^{\rm 1PN}_{\text{AT}} + E^{\rm 2PN}_{\text{AT}} + E^{\rm 3PN}_{\text{AT}} \,,
\end{align}
where the energy up to 2PN is presented in \cite{Mandal:2023lgy} and the 3PN result is given as,
\begin{align}
E^{\rm 3PN}_{\text{AT}} &=  x^9 \left\{ \left[-
\frac{45 }{32}\nu ^3+\frac{74495 }{448}\nu ^2+  \left(-\frac{823243}{784}+\frac{8895 \pi ^2}{512}\right)\nu +\frac{894721}{3136}
+ \frac{321}{7} \left( 2\nu-1 \right) \log\left(x\widetilde{R}_{\rm orb}\right)  \right] \widetilde{\lambda}_{(+)} \nonumber\right.\\
&\quad\quad\quad\quad+ \left. \left[
\frac{825 }{64}\nu ^2-\frac{42225 }{224} \nu +\frac{378751}{3136} - \frac{321}{7} \log\left(x\widetilde{R}_{\rm orb} \right) \right]\delta \widetilde{\lambda}_{(-)} - 45 (\widetilde{\lambda}\kappa)_{(+)}   \right\} \,,
\end{align}
where $x=\widetilde{\omega}^{2/3}$, $\widetilde{R}_{\rm orb}=c^2 R_{\rm orb}/(G_N M)$ and the combined dimensionless Love numbers are denoted by
\begin{align}
\label{eq_repair_1}
\widetilde{\lambda}_{(\pm)} &=
\frac{m_{(2)}}{m_{(1)}} \widetilde{\lambda}_{(1)}
\,\pm\,
\frac{m_{(1)}}{m_{(2)}} \widetilde{\lambda}_{(2)} \,, \\
(\widetilde{\lambda}\kappa)_{(\pm)} &=
\widetilde{\lambda}_{(1)} \kappa_{(1)}(R_{\rm orb})
\,\pm\,
\widetilde{\lambda}_{(2)} \kappa_{(2)}(R_{\rm orb}) \,.
\end{align}
Here, $\kappa_{(a)}(R_{\rm orb})$ is evaluated at the orbital length scale $R_{\rm orb}$, after evolving it using the RG Eq. \eqref{eq_RG_sol} from the matching scale $R_{\rm NS}$. 

\subsection{Scattering angle}
\label{sec_scattering_angle}
Here we present the scattering angle $\chi$ in the COM frame for the hyperbolic encounter of two stars. 
We begin by expressing the Hamiltonian $\mathcal{H}$ (which is a function of $p_r$, $L$ and $r$) to obtain $p_r = p_r(\mathcal{H},L,r)$.
Then, we use relation between the Lorentz factor $\gamma$ and the total energy per total rest mass $\Gamma = \mathcal{H}/(Mc^2)$ given by 
\begin{align}\label{eq_gamma_for_v}
\gamma=\frac{1}{\sqrt{1-v^2/c^2}}=1+\frac{\Gamma^2-1}{2\nu}\, ,
\end{align}
where $v\equiv|\dot{\bm{r}}|$ is the relative velocity of the compact objects, 
and the total angular momentum $L$ and the impact parameter $b$ are related by
$L=(\mu\gamma v b)/\Gamma$.
This allows us to exchange $\mathcal{H}$ for $v$ and $L$ for $b$. 
With this, we can then write the scattering angle as
\begin{align}
\chi(v,b) = - \frac{\gamma}{\mu\gamma v} \, \int \dd r \, \frac{\partial p_r(v,b,r)}{\partial b}  - \pi\, .
\end{align}

Performing this procedure with the Hamiltonian~\eqref{eq_ham_ad} yields the scattering angle computed in the COM frame, which we write as
\begin{align}
   \chi_{\rm AT} =\chi^{\rm 0PN}_{\rm AT}+\chi^{\rm 1PN}_{\rm AT}+\chi^{\rm 2PN}_{\rm AT}+ \chi^{\rm 3PN}_{\rm AT} 
\end{align}
where the scattering angle up to 2PN is presented in \cite{Mandal:2023lgy} and the 3PN result is given as
\footnote{For computing the logarithmic terms in the scattering angle, check Ref.~\cite{Bini:2017wfr}.}
\begin{align}
\label{eq:SA_ES2_N3LO}
\frac{\chi^{\rm 3PN}_{\rm AT}}{\Gamma} &=  \frac{v^2}{M b^4}
\begin{bmatrix}%
\lambda_{\rm (+)} & ~\delta \lambda_{\rm (-)}
\end{bmatrix}
\cdot \left\{  
\pi \left(\frac{G_NM}{v^2b}\right)^2 \frac{1575}{256} \begin{bmatrix} 1\\ 0\end{bmatrix} \left(\frac{v^6}{c^6}\right)   + \left(\frac{G_NM}{v^2b}\right)^3 \frac{1}{70}  \begin{bmatrix} 18073\\2713 \end{bmatrix}  \left(\frac{v^6}{c^6}\right) \right. 
\nonumber\\
&\quad \left. +\, \pi \, \left(\frac{G_NM}{v^2b}\right)^4 \left( \frac{1}{458752}\begin{bmatrix} 304535296-\left(46848512+231525 \pi ^2\right) \nu\\ 51930496\end{bmatrix} 
+ \frac{1605}{64} \log\left(\frac{2b}{R_{\rm sc}}\right)\begin{bmatrix} 1-2\nu\\ 1\end{bmatrix}
\right) \left(\frac{v^6}{c^6}\right) \right.
\nonumber\\
&\quad\left. +\, \left(\frac{G_NM}{v^2b}\right)^5 \left\{ 192\begin{bmatrix} 1\\ 0\end{bmatrix} +48\begin{bmatrix}53-4\nu\\ 5\end{bmatrix}  \left(\frac{v^2}{c^2}\right) + \frac{12}{35} \begin{bmatrix} 28753-3480\nu \\ 7473-140\nu\end{bmatrix}  \left(\frac{v^4}{c^4}\right) \right.\right.
\nonumber\\
&\quad\quad\quad\quad\left.
+ \left(\frac{1}{8575} \begin{bmatrix}\left(584325 \pi ^2-68190952\right) \nu +128915306 \\ 33894506-2486400 \nu\end{bmatrix} 
+ \frac{903936}{1225} \log\left(\frac{b}{2R_{\rm sc}}\right)\begin{bmatrix} 1-2\nu\\ 1\end{bmatrix} 
\right) \left(\frac{v^6}{c^6}\right) \right\} \Bigg\} 
\nonumber\\
&\quad+\frac{v^2}{M b^4}
\begin{bmatrix}%
(\lambda\kappa)_{\rm (+)} & ~\delta (\lambda\kappa)_{\rm (-)}
\end{bmatrix}
\cdot \left\{  
 \pi \, \left(\frac{G_NM}{v^2b}\right)^4 \frac{1575}{64} \begin{bmatrix} -\nu\\ 0\end{bmatrix} \left(\frac{v^6}{c^6}\right) +\, \left(\frac{G_NM}{v^2b}\right)^5 \frac{25344}{35}\begin{bmatrix} -\nu\\ 0\end{bmatrix} 
 \left(\frac{v^6}{c^6}\right)  \right\} 
\nonumber\\
&\quad +\mathcal{O}\left(G_N^6,\frac{v^8}{c^8}\right) \,,
\end{align}

where combined Love numbers are denoted by,
\begin{align}
{\lambda}_{(\pm)} &=
\frac{m_{(2)}}{m_{(1)}} {\lambda}_{(1)}
\,\pm\,
\frac{m_{(1)}}{m_{(2)}} {\lambda}_{(2)} \,,\\
\label{eq_repair_2}
({\lambda}\kappa)_{(\pm)}&=
 {\lambda}_{(1)} \kappa_{(1)}(R_{\rm sc})
\,\pm\,
 {\lambda}_{(2)} \kappa_{(2)}(R_{\rm sc})\,.
\end{align}
Here, $\kappa_{(a)}(R_{\rm sc})$ is evaluated at the typical length scale $R_{\rm sc}$ of the scattering system, after evolving it using the RG Eq. \eqref{eq_RG_sol} from the matching scale $R_{\rm NS}$. 
Equation~\eqref{eq:SA_ES2_N3LO} agrees to 3PM (i.e.,~$G_N^3$) with the result reported by Ref.~\cite{Jakobsen:2022psy,Kalin:2020lmz}, obtained using techniques of worldline QFT~\cite{Mogull:2020sak} and the EFT developed for PM calculations~\cite{Kalin:2020mvi}, respectively.
\section{Conclusions}\label{sec_Conclusion}
 
We calculated the conservative two-body effective Hamiltonian, taking into account the dynamical tidal interaction up to 3PN order.
Our approach involves Feynman diagrams up to three loops, which are evaluated using the dimensional regularization scheme. 
We also studied the adiabatic limit and obtained the adiabatic tidal Hamiltonian up to 3PN order. 
The fulfillment of Poincar\'e algebra constituted an important validation of the Hamiltonian. 
Additionally, we presented the analytic expressions of two gauge-invariant observables, namely the binding energy for bound orbits and the scattering angle for hyperbolic encounters.

In the considered action, we included the contribution from the non-minimal coupling (last term of Eq.~\eqref{eq_Lag_intro}), which turned out to be crucial for ensuring the finite observables at the 3PN order. Interestingly, we found that not all the divergent pieces can be removed by adding a total derivative to the Lagrangian (canonical transformations on the Hamiltonian). Therefore, we constructed a counter-term to reabsorb those divergent pieces, yielding the renormalization of the 
the post-adiabatic Love number $\kappa_{(a)}$. Consequently, the latter experiences a renormalization group running, as shown in Sec.~\ref{sec_renorm_pAdterm}. According to the corresponding beta function, we observed that the value of $\kappa_{(a)}$ increases, when reducing the external length scale $R$.

In the adiabatic limit, our results depend on the two Love numbers (Wilson coefficients within the context of point particle EFT), namely, $\lambda$ (the tidal Love number) and $\kappa$ (the post-adiabatic Love number). 
The determination of their specific values in the case of BHs requires a matching computation of an observable between the EFT and BHPT, as described  in~\cite{Kol:2011vg,Chia:2020yla,Hui:2020xxx,Charalambous:2021mea,Hui:2021vcv,Ivanov:2022qqt,Ivanov:2022hlo}. Notably, the tidal Love number $\lambda$ is found to be zero for BHs~\cite{Damour:1982wm,damour_1984_love_number,Kol:2011vg,Damour:2009vw,Binnington:2009bb}. However, the specific value of the post-adiabatic Love number $\kappa$ remains unknown and requires further investigation. 
In the case of NSs, Love numbers are typically expressed in terms of their EOS. For $\lambda$, this was accomplished in Refs.~\cite{Hinderer:2007mb,Flanagan:2007ix}, and $\kappa$ in Ref.~\cite{Chakrabarti:2013lua}.
It will be interesting to compute the particular values of $\kappa$ for different compact objects~\cite{Cardoso:2017cfl,Brustein:2020tpg,Raposo:2018rjn,Nair:2022xfm}. The implications of the flow of the post-adiabatic Love number on the EOS of NS are very intriguing and warrant further investigation.
An interesting implication of the running post-adiabatic Love number is that strictly adiabatic tides are impossible starting from the 3PN order of accuracy, but instead tides have to be dynamical.
This also makes sense from a microscopic point of view: An exactly instantaneous tidal response is impossible since the propagation speed within the body cannot exceed the speed of light.

The adiabatic Love number $\lambda$ was found to have a non-zero value for Schwarzschild BHs in spacetime dimensions other than four, as shown in Ref.~\cite{Kol:2011vg} \footnote{For analysis of Love numbers when nonlinearities in the theory are taken into account, see Ref.~\cite{DeLuca:2023mio}.}. Consequently, when using dimensional regularization, the Love number $\lambda$ yields a finite contribution at $\mathcal{O}(\epsilon=d-3)$, which, in combination with the divergent terms generated by the $\mathcal{L}_{\rm EQ}$, results in a finite $\mathcal{O}(\epsilon^0)$ contribution. 
However, such contributions to observables could be absorbed into the contributions from the counterterm, 
so that the number of degrees of freedom to be matched does not increase.
Furthermore, matching in arbitrary dimension $d$ or in $d=3$ will in general result in different values for $\lambda$ and $\kappa$, but observables are the same in the limit $\epsilon \rightarrow 0$.
We note that a matching for generic $d$ might require the inclusion of evanescent operators \cite{Collins:1984xc}, 
which are non-zero only in spacetime dimensions other than four. 
Indeed,
it was observed in \cite{Dugan:1990df, Buras:1989xd, Herrlich:1994kh}, that the scattering amplitudes from such operators can affect the matching equation determining the Wilson coefficients. Also the appearance of such operators in counterterms originating from physical operators leads to the mixing of physical and evanescent operators during the RG evolution~\cite{Dugan:1990df}.

In this work, we focused on the dynamical gravitoelectric quadrupolar tides, but our framework and our automatic computational techniques are general enough to be extended to obtain further higher-order corrections and also other types of tidal effects, like the dynamical gravitomagnetic tides~\cite{Gupta:2020lnv,Gupta:2023oyy}, as well as to incorporate higher order multipolar tides.
The coupling of the oscillation modes of the NS with other degrees of freedom, such as its spin~\cite{Ho:1998hq,Steinhoff:2021dsn,Ma:2020rak,Bhatt:2023zsy}, or other oscillation modes~\cite{2012ApJ...751..136W,2013ApJ...769..121W,Landry:2015snx,Xu:2017hqo,Essick:2016tkn,Yu:2022fzw} could also be incorporated.
Finally, our 3PN Hamiltonians can be added to the time-domain effective-one-body waveform models of Refs.~\cite{Hinderer:2016eia,Steinhoff:2016rfi}.

\subsection*{Acknowledgements}

We thank Sumanta Chakraborty, Thibault Damour, Rossella Gamba, Quentin Henry, Mikhail Ivanov, Gustav Jakobsen, Jung-Wook Kim, Oliver Long, Elisa Maggio, Saketh MVS, Alessandro Nagar, Ira Rothstein and Justin Vines for insightful comments and discussions.
%
We would like to thank authors of \cite{Jakobsen:2023pvx} for pointing out a typo in equation \eqref{eq_repair_2}.
%
The work of M.K.M is supported by Fellini~-~Fellowship for Innovation at INFN funded by the European Union's Horizon 2020 research and innovation programme under the Marie Sk{\l}odowska-Curie grant agreement No.~754496. 
%
%
H.O.S acknowledges funding from the Deutsche Forschungsgemeinschaft (DFG)~-~Project No.~386119226.
R.P.'s research is funded by the Deutsche Forschungsgemeinschaft (DFG, German Research Foundation), Projektnummer 417533893/GRK2575 “Rethinking Quantum Field Theory”.

\appendix

\bibliographystyle{JHEP}
\bibliography{biblio}

\end{document}